\documentclass[aps,prb,floatfix,twocolumn,superscriptaddress,showpacs,longbibliography,epsf]{revtex4-1}

\usepackage[USenglish]{babel}
\usepackage{amsmath,amssymb,amsfonts}
\usepackage{epstopdf}
\usepackage{epsfig} 
\usepackage{graphicx}
\usepackage{subfigure}
\usepackage{import}
\usepackage{color}
\usepackage{url}
\usepackage[final]{changes}
\allowdisplaybreaks

\usepackage[colorlinks=true,linkcolor=blue,citecolor=red]{hyperref}

\makeatletter
\renewcommand\tableofcontents{\@starttoc{toc}}
\makeatother
\def\bcen{\begin{center}}
\def\ecen{\end{center}}

\def\a{\alpha}       \def\b{\beta}   \def\g{\gamma}   \def\d{\delta} 
\def\e{\varepsilon}          
       \def\l{\lambda} \def\m{\mu}      \def\n{\nu}
                    \def\s{\sigma}
\def\t{\tau}           
        \def\o{\omega}   
\def\G{\Gamma}       \def\D{\Delta}      
                     
        \def\O{\Omega}

\def\aa{{\V \a}}

\def\eg{\mbox{\it e.g.\ }}  \def\ie{\mbox{\it i.e.\ }}
\def\=={\equiv}

\def\qed{\raise1pt\hbox{\vrule height5pt width5pt depth0pt}}

\def\cG0{{\cal G}_0} 
\def\cG{{\cal G}}

\def\bk{{\bf k}}
\def\bq{{\bf q}}
\def\bp{{\bf p}}

\def\ie{\hbox{\it i.e.\ }} \def\eg{\mbox{\it e.g.\ }}
\def\ie{\mbox{\it i.e.\ }} \def\=={\equiv}

 \def\ep0{\epsilon_{p}} \def\ed0{\epsilon_{f}}

\def\be{\begin{equation}}
\def\ee{\end{equation}}

\def\cc{c^{\dagger}}
\def\ca{c^{\phantom{\dagger}}}

\def\ac{a^{\dagger}}
\def\aa{a^{\phantom{\dagger}}}

\newcommand{\bd}[1]{\mathbf{#1}}
\newcommand{\bs}[1]{\boldsymbol{#1}}

\newcommand{\quave}[1]{\langle{#1}\rangle}

\def\bx{\mathbf{x}}

\usepackage{fancyhdr}
\begin{document}

\author{Ivan Amelio}
\affiliation{
Institute of Quantum Electronics ETH Zurich, CH-8093 Zurich, Switzerland}
\affiliation{INO-CNR BEC Center and Dipartimento di Fisica, Universit{\`a} di Trento, 38123 Povo, Italy}

\author{Lukas Korosec}
\affiliation{DQMP, Universit{\'e} de Gen{\`e}ve, 24 quai Ernest Ansermet, CH-1211 Gen{\`e}ve, Switzerland}

\author{Iacopo Carusotto}
\affiliation{INO-CNR BEC Center and Dipartimento di Fisica, Universit{\`a} di Trento, 38123 Povo, Italy}

\author{Giacomo Mazza}
\email{giacomo.mazza@unige.ch}
\affiliation{DQMP, Universit{\'e} de Gen{\`e}ve, 24 quai Ernest Ansermet, CH-1211 Gen{\`e}ve, Switzerland}

\title{Optical dressing of the electronic response of two-dimensional semiconductors 
in quantum and classical descriptions  of cavity electrodynamics}

\begin{abstract}
\added{We study quantum effects of the vacuum light-matter interaction 
in materials embedded in optical cavities.}
\replaced{We focus}{We study effects of light-matter hybridization} on the electronic 
response of a two-dimensional semiconductor placed 
inside a planar cavity.
By using a diagrammatic expansion of the electron-photon 
interaction, we describe signatures of light-matter hybridization
characterized by large asymmetric shifts of the spectral weight 
at resonant  frequencies.
We follow the evolution of the light-dressing from 
the cavity to the free-space limit.
In the cavity limit, light-matter hybridization
results in a  modification of the optical gap
with sizeable spectral weight appearing below the bare gap edge.
In the limit of large cavities, we find a residual redistribution of spectral weight
which becomes independent of the distance between the two mirrors.
We show that the photon dressing of the electronic response
can be fully explained by using a classical description of light. 
The classical description is found to hold up to a 
strong coupling regime of the light-matter interaction highlighted
by the large modification of the photon 
spectra with respect to the empty cavity.
\deleted{We assess the importance of corrections due to
the quantum nature of light. }
We show that, despite the strong coupling, quantum corrections 
are negligibly small and weakly dependent on the cavity confinement.
As a consequence, in contrast to the optical gap, the 
single particle electronic band gap 
is not sensibly modified by the strong-coupling.
\added{Our results show that 
quantum corrections are dominated by off-resonant photon 
modes at high energy.
As such, cavity confinement can hardly 
be seen as a knob to control the quantum 
effects of the light-matter interaction in vacuum.}
\end{abstract}

\maketitle

\section{Introduction}
The interaction of light and matter in vacuum conditions, \ie 
in the absence of driving fields, 
has been proposed as an advanced frontier 
for the control of matter properties.~\cite{ebbesen_perspective}
In this perspective, electromagnetic environments
capable of confining light over small volumes, namely cavities, 
play a fundamental role.

The spatial confinement can enhance fluctuations of electromagnetic fields 
and in turn result in a significant increase of the strength of the vacuum 
light-matter interaction.
These effects become manifest in the so-called strong and ultra-strong 
coupling regimes of light-matter interaction in which matter and light 
degrees of freedom mix 
and form strongly intertwined excitations
known as polaritons.~\cite{nori_review,ciuti_carusotto_RMP}

Recently, several proposals have suggested  
light-matter hybridization as a novel tool to 
 control emergent macroscopic phenomena, 
such as superconductivity, 
ferroelectricity~\replaced{and magnetic or topological phases}{ or spin liquids}
by embedding so-called quantum materials in such 
electromagnetic environments.~\cite{sentef_superconductivity,schlawin_superconductivity,
curtis_superconductivity_cavity,PhysRevX.10.041027,chiocchetta2020cavityinduced,
ebbesen_superc,debernardis2018,schuler2020,
colombelli_exciton_bounds,andolina_spatially_varying,
guerci_superradiance_2020,nataf_rashba,wang_chern,
rubio_nature_material_chiral}
\added{The general idea is that when a material is embedded inside a cavity, 
interaction effects due to vacuum quantum fluctuations of the electromagnetic 
fields become strong enough to induce major modifications 
of the ground state properties of the material.

}
The new perspective has been termed 
the quantum route towards  the light-manipulation 
of matter,~\cite{kiffner_quantum_light,jiajun_martin} 
as opposed to the  classical approach which instead 
relies on the 
stimulation of matter by means of coherent driving fields.~\cite{nicoletti_review,non_linear_phononics_ybco,mitrano_K3C60} 
The main difference between the two approaches can be traced back 
to the different roles played by the photonic degrees 
of freedom in the dynamics of the coupled light-matter system.
In the classical regime, the driving field contains 
a macroscopically large number of photons
whose dynamics is assumed to be entirely determined 
by external sources.
In contrast, in the quantum case 
vacuum photons are considered to be active degrees 
of freedom whose dynamics is self-consistently 
determined by the  coupling with the matter 
microscopic excitations
as described in the framework of the 
non-relativistic Quantum Electrodynamics (QED).

Despite this conceptual distinction, 
the differences between the two regimes 
of the light-matter interaction can be, in reality, 
much less sharply defined.
Indeed, as originally pointed out in a seminal work by 
Jaynes and Cummings,~\cite{jc_original} 
semiclassical approaches 
based on Maxwell's equations can be extremely 
effective  in describing light-matter coupling 
down to  the vacuum limit. 
At the same time, even in the presence of coherent 
driving fields, the feedback of internal sources onto 
driving photons can play an important role
as recently discussed in the context
of  light-stimulated strongly correlated 
electron systems.~\cite{michele_minimal_coupling}

\replaced{In this context,}{As new proposals emerge,~\cite{rubio_nature_material_chiral}}
a deeper understanding  of the crossover between quantum and classical regimes
of the light-matter interactions as a function of the 
environment represents a fundamental step towards
practical cavity applications of solid-state materials.~\cite{YLaplace_cavity}
General open questions concern, for example, 
the effectiveness of the coupling in the modification of the 
ground state properties of a material and how these effects depend
on the resonance conditions between multiple matter 
excitations and the electromagnetic modes of the environment.

In this paper, we address these questions by analysing the 
effects of light-matter hybridization on the electronic 
response of a two dimensional semiconductor embedded in a planar cavity. 
Photon dressing of the dc electronic transport properties has been recently 
discussed in the context of organic semiconductors~\cite{orgiu_organic_semiconductor,
hagenmuller_cavity_transport,ebbesen_acs_nano_2020}
and two-dimensional electron 
gases.~\cite{rokaj2021free,bartolo_ciuti_magnetotrasport,feist_magnetostransport}
These descriptions usually rely on effective models in which 
the effect of the cavity is incorporated as effective coupling strengths.
Here we consider a multi-mode electron-photon 
Hamiltonian for which, at fixed electronic structure of the material, 
the light-matter interaction is entirely controlled by the 
distance  between the mirrors.
This allows us to 
follow the evolution of the light-dressing 
from the  {\it cavity limit}, characterized by a strong light confinement, 
to the {\it free-space limit}, in which effects 
of the light-matter interaction become independent 
of the environment.

By making use of a diagrammatic expansion of the electron-photon 
interaction, we show  sharp redistributions  of spectral weight 
in the frequency dependent conductivity due to the hybridization 
of resonant modes with the continuum of electronic excitations 
in the material.
The shift results in a modification of the optical gap
which smoothly evolves into a residual sub-gap redistribution 
of the spectral weight persisting up to the free-space limit. 
We rationalize the results of the diagrammatic expansion 
in terms of a semi-classical description of the vacuum fluctuations 
in which classical electromagnetic fields are sourced by current  
fluctuations  in the material. 
By comparing the two approaches we show that the classical 
description exactly reproduces  the results of the  diagrammatic 
expansion in the Gaussian approximation up to the strong coupling limit.

We estimate the size of quantum effects by computing 
corrections beyond  the Gaussian approximation. We show 
that, because of the small photon density of states at 
low energies, quantum corrections are expected to be 
negligibly small and~\added{dominated by off-resonant modes 
at high energy. Therefore, quantum corrections are}{ weakly dependent 
on the cavity confinement.}
The direct consequence of our results is that, while the 
optical gap, described by the current-current response function,  
is sensibly modified by the optical dressing
in the cavity, the electronic band gap, determined by the 
single-particle electronic Greens' function, remains unchanged.

We organize the paper as follows.
In section~\ref{sec:framework} we present
the setup studied in this work. 
Section~\ref{sec:linear_response} reports the 
detail of the linear response theory and the 
diagrammatic expansion. 
In section~\ref{sec:results_transport} we present main 
results concerning signatures of the light-matter hybridization 
in the conductivity. Eventually, in section~\ref{sec:classical_quantum} 
we address the comparison between the classical description and 
the quantum theory.

\section{Theoretical Framework}
\label{sec:framework}
The theoretical framework used in this work 
is specified by the cavity,  
the electronic system, and the interaction Hamiltonian.

\begin{figure}
\includegraphics[width=\columnwidth]{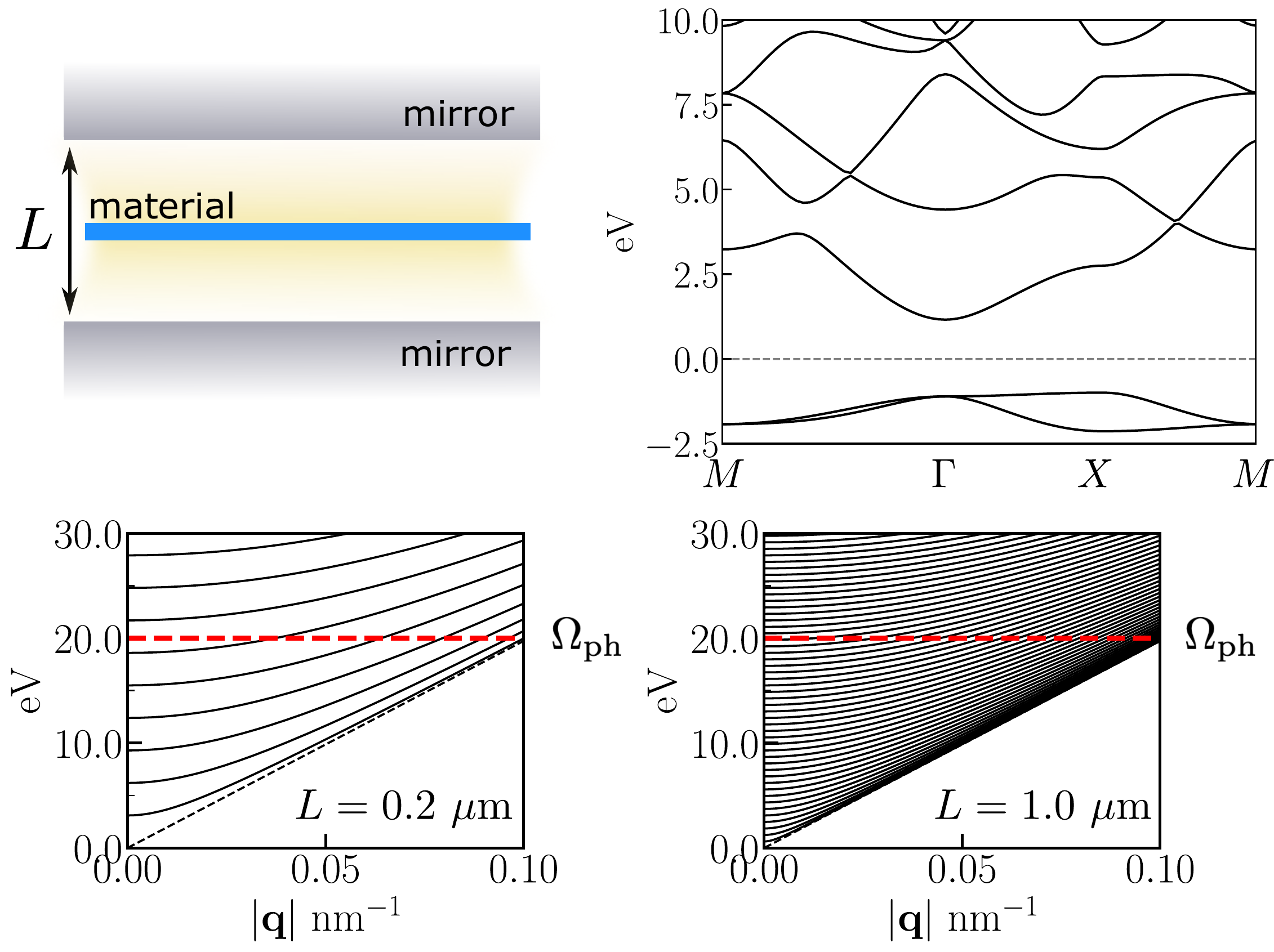}
\caption{Top: (left) Sketch of a two-dimensional
material placed between two perfectly conducting mirrors separated by a distance $L$.
(right) Electronic bands of the two-dimensional system
as obtained by using the periodic potential defined by the parameters
$v_0 = 5.0~\mathrm{eV}$, $\l=0.2~\mathrm{nm}^{-1}$, $\a=2$
and $\xi=0.1~\mathrm{nm},$ see text.
Bottom: 
Photon dispersions for two cavity lengths. 
Red dashed line indicate a photon cutoff used in the calculations.
The massless mode drawn with a dashed line represents 
the mode with zero in-plane component.
}
\label{fig:fig1}
\end{figure}

\subsection{Cavity}
We consider a two-dimensional material placed in the center of a 
planar cavity made by two parallel perfectly conducting mirrors
separated by a distance $L$, see Fig.~\ref{fig:fig1}.
Following  Ref.~\onlinecite{kakazu_quantum_cavities}  the quantization of the 
electromagnetic field in the planar geometry  leads to the expansion of the 
vector potential operator 
\begin{equation}
\bd{A}(\bd{x},z) = \sum_{\bq q_z \sigma} \frac{\g_{\bq q_z}}{\sqrt{V}}
\left( 
\bs{w}_{\bq q_z \s}(\bd{x},z) \ac_{\bq q_z \sigma} + 
\mathrm{h.c.}
\right)
\end{equation}
where $\ac_{\bq q_z \s}$ are creation operators 
of photons with in-plane momentum $\bq$, 
out-of-plane
momentum $q_z$ and polarization $\s$.
The amplitudes 
$\g_{\bq q_z} \equiv \sqrt{\frac{\hbar^2}{2 \epsilon_0 \omega_{\bq q_z}}}$
depends on the energy of the photon modes
$ \o_{\bq q_z} \equiv  \hbar c \sqrt{\bq^2 + q_z^2}$.
In the rest of the paper we will measure all the frequencies in energy units
by incorporating the $\hbar$ in the definition of frequency.
$V = S \times L$ is the cavity volume, being $S$
the area of the mirrors, which is assumed to be equal to that
of the embedded material. 
Eventually we will take the thermodynamic
limit, by sending $S \to \infty$ and keeping the density 
of electrons  in the material fixed.
As a consequence  $\bq$ is described by a 
continuous variable in the $(q_x,q_y)$ plane, 
whereas $q_z$ is quantized  as $q_z = \frac{\pi}{L}n$, with $n$ 
a positive integer.
Dispersion of the modes for two cavity lengths are reported in Fig.~\ref{fig:fig1}. 

The mode functions $\bs{w}_{\bq q_z \s}(\bd{x},z) 
\equiv e^{i \bq \bx} \bs{v}_{\bq q_z \s}(z)$ are determined by imposing the 
boundary conditions at the mirrors $\bd{E}_\parallel(\bd{x},z=0) = \bd{E}_\parallel(\bd{x},z=L) = 0$, $\bd{B}_\perp(\bd{x},z=0) = \bd{B}_\perp(\bd{x},z=L)=0 $ and the transversality condition
$\nabla \cdot \bd{A} = 0.$
For each momentum $(\bq,~q_z)$ there exist two orthogonal 
polarizations $\bs{w}_{\bq q_z \s} \cdot \bs{w}_{\bq q_z \s'} \propto \d_{\s \s'}.$
The mode functions are specified by the vectors 
$ \bs{v}_{\bq q_z \s}(z)$ which depend on 
the momenta and polarization.~\cite{kakazu_quantum_cavities}

\subsection{Electrons}
We model the electronic properties of the material by considering 
a two-dimensional periodic potential on a square lattice 
with lattice parameter $a=0.5~\mathrm{nm}$. 
The periodic potential is defined by $V_\mathrm{periodic}(\bd{x}) = \sum_{\bd{R}} v(\bd{x} -\bd{R})$
where the sum extends over all the points in the Bravais lattice and 
$v(\bd{x})$ represents a deep potential well centred in the middle 
of the square unit cell mimicking an atomic center.
We define the atomic-like potential as
$v(\bd{x}) = -v_0 \exp(-\l |\bx|)/(|\bx|^\a+\xi^\a)$. The parameters
$v_0$, $\l$, $\a$ and $\eta$ are chosen in order to get a
band structure with bandwidths of the order 
of the $\mathrm{eV}$, see Fig.~\ref{fig:fig1}.

We tune the chemical potential in the middle of the band gap.
The size of the indirect band gap is $\sim 2 ~\mathrm{eV}$,
slightly smaller than the direct optical gap at the $\G-$point 
$\sim 2.25~\mathrm{eV}$.
The details of the band structure, as for example the fact that the electronic gap 
is slightly indirect, do not affect in any qualitative way 
the results presented in this paper.

The main goal of this paper is to estimate the effects of
hybridization of matter and transverse photons. 
As such we do not consider the effects of the electron-electron 
interactions as well as effects on the screening of the 
Coulomb interaction due to the 
image charges on the mirrors.\cite{todorov_intersubband_polaritons,debernardis2018}
Throughout the paper the electrons-photons interaction represents 
the only source of scattering.
We also consider the electrons spinless.

\subsection{Multi-mode QED Hamiltionian}
\label{sec:mulit_mode_QED_Hamiltonian}
The electron-photon interaction is introduced by the 
minimal coupling $\bp \to \bp + e \bd{A}(\bx)$, 
\begin{equation}
\begin{split}
H =& H_\mathrm{0} + 
\int d\bx~\Psi^\dagger(\bx) \frac{\left( \bp + e \bd{A}(\bx) \right)^2}{2 m } \Psi(\bx) \\
 &+  \int d\bx~\Psi^\dagger(\bx) V_\mathrm{periodic}(\bx) \Psi(\bx) 
\end{split}
\label{draft_eq:Hgeneral}
\end{equation}
where 
$\bd{A}(\bx) \equiv \bd{A}(\bx,z=L/2)$
is the vector potential operator evaluated in the middle 
of the cavity.
The fermionic fields $\Psi(\bx) = 
\sum_{\bk \n}  \varphi_{\bk \n}(\bx) \ca_{\bk \nu}$ 
are expanded onto the complete set of Bloch wavefunctions 
$\varphi_{\bk \n} (\bx)$ of the two-dimensional periodic potential.
$H_0 = \frac{1}{2}\int d \bd{x} d z \ \e_0 \bd{E}^2(\bd{x},z) +\frac{1}{\mu_0} \bd{B}^2(\bd{x},z)  
= \sum_{Q}  \o_Q \ac_Q \aa_Q$ is the total energy of the electromagnetic field inside the 
cavity. In the last expansion we used the compact notation $Q \equiv (\bq,q_z,\s)$.
Expansion of the Hamiltonian gives
\begin{equation}
H= \sum_Q \o_Q \ac_Q \aa_Q + \sum_{\bk \n} \e_{\bk \n} \cc_{\bk \n} \ca_{\bk \n}
+ H_{AP} + H_{AA}
\end{equation}
with
\begin{equation}
H_{AP} = 
\frac{e}{m} \sum_{Q} \frac{\g_{Q}}{\sqrt{V}} \left( \bs{v}_{Q} \cdot \bd{P}(\bq) \aa_Q + h.c. \right)
\label{draft_eq:H_AP}
\end{equation}
\begin{equation}
\begin{split}
H_{AA} = 
\frac{e^2}{2 m}\sum_{Q Q'} \frac{\g_Q}{\sqrt{V}}  \frac{\g_{Q'}}{\sqrt{V}}
\left( \bs{v}^*_Q \cdot \bs{v}_{Q'} \hat{\rho}(\bq'-\bq) \ac_{Q} \aa_{Q'} +  \right. \\
\left.  \bs{v}^*_Q \cdot \bs{v}^*_{Q'} \hat{\rho}(-\bq'-\bq) \ac_{Q} \ac_{Q'} + h.c. \right)
\end{split}
\label{draft_eq:H_AA}
\end{equation}
In the last equations $\bd{P}(\bq) \equiv \int d\bx 
\Psi^{\dagger}(\bx) e^{i \bq \bx} \bd{p} \Psi (\bx)$,
$\hat{\rho}(\bq) \equiv \int d \bx \Psi^{\dagger}(\bd{x}) e^{i \bq \bd{x}} \Psi(\bd{x}) $	
are, respectively, the momentum and density operators 
at in-plane wavevectors $\bq$
and
$\bs{v}_Q \equiv \bs{v}_Q(z=L/2)$.

The coupling constants that define the interaction terms $H_{AP}$
and $H_{AA}$ depend on the electronic structure of the material
through the matrix elements 
of the momentum and density operators
and on the cavity geometry through the mode functions $\bs{v}_Q$ 
and the amplitudes $\g_Q$.
We keep the electronic structure fixed
so that the light-matter interaction is controlled 
by the distance between the mirrors $L$.

As $L$ is increased, the photon spectrum
becomes denser, as shown in Fig.~\ref{fig:fig1}.
Here, in order to properly take into account 
the effects of the light-matter interaction in both 
limits of small and large cavities, 
we retain the full multi-mode structure of the photon field.
We notice that, in principle, the multi-mode 
QED Hamiltonian, Eq.~\ref{draft_eq:Hgeneral},
describes interaction with photon modes at all energy scales.
However, as we are interested in interaction  
effects due to coupling with low-energy photons 
confined by the mirrors, we cutoff the photon 
spectrum at an energy $\O_\mathrm{ph}$.
Physically, this cutoff corresponds 
to the largest energy of photon modes that 
can be confined by the cavity
and is expected to be of the order of the  
plasma frequency of the mirrors.
We assume all the effects of renormalization of the 
electronic  properties due to interaction with 
photons of energy higher than the cutoff to be already 
included in the electronic  structure 
defined by the periodic potential $V_\mathrm{periodic}(\bx)$.
If not explicitly stated otherwise the photon spectrum 
cutoff is set to $\Omega_\mathrm{ph}=20~\mathrm{eV}$.

In order to ensure invariance of the results with respect to 
equivalent  representations of the light-matter Hamiltonian
we expand the electronic Hamiltonian 
onto a large  subspace of electronic degrees of freedom.~\cite{debernardis_breakdown,
nori_representations_nat_phys,schuler_deveraux_2021,
jiajun_tight_binding,ashida2021,schiro_dmytruk_2021} 
Specifically, we retain a total of $N_b=30$ electronic bands.

All the presented results are converged with respect to the  
cutoff in the number of electronic bands. 
Convergence with respect to the photon cutoff 
will be discussed in the rest of the paper.

\section{Linear Response Theory}
\label{sec:linear_response}
We investigate the effects of the light-matter interaction
on the  material properties by focusing on the 
long wavelength response to an arbitrary small electric field.
The electronic response is defined by the conductivity tensor
\begin{equation}
\sigma_{ij}(\o) = 
\frac{\chi_{ij}(\o)}{i \o} + \frac{\d_{ij}}{i \o} \frac{e^2}{m} \rho_0
\label{draft_eq:sigma_def}
\end{equation}
where $\rho_0$ is the average electronic density 
and
\begin{equation}
\chi_{ij}(t) = -i \theta (t) \quave{\lbrace J_i (t), J_j \rbrace}
\end{equation}
is the current-current response function
being $\bd{J} = \frac{1}{S}\int d\bx \bs{j}(\bx)$
the average current density.
Here we consider a system with time reversal symmetry 
so that $\s_{ij} = \d_{ij} \s_{ii}$. 
Moreover, due to the symmetry of the square lattice
$\s_{xx} = \s_{yy} =\s.$

Since the electronic problem is strictly two-dimensional
$\bs{j}(\bx)$ represents a surface current density
and is defined by the continuity equation 
$i e [\Psi^\dagger \Psi , H] = -\nabla \bs{j}$. The  
current density operator contains a purely electronic term 
plus a  diamagnetic contribution coming from the vacuum 
electromagnetic field $\bs{j}(\bd{x}) = \bs{j}_P(\bx) + \bs{j}_A(\bx)
$
\begin{equation}
\bs{j}_P(\bd{x}) = 
\frac{e}{2m} \Psi^\dagger(\bx) \bp \Psi(\bx)+ h.c.
\label{draft_eq:jp_def}
\end{equation}
\begin{equation}
\bs{j}_A(\bd{x}) = 
\frac{e^2}{m} \Psi^\dagger(\bx) \Psi(\bx) \bd{A}(\bx)
\label{draft_eq:jp_def}
\end{equation}
The response function decomposes in three  
terms $\chi_{JJ} = \chi_{PP} + \chi_{PA} + \chi_{AA}$
defined, respectively, as the purely electronic 
$\chi_{PP} = \chi_{J_P,J_P}$, 
the mixed 
$\chi_{PA} = \chi_{J_P,J_A} + \chi_{J_A,J_P} $ and
the purely diamagnetic $\chi_{AA} = \chi_{J_A,J_A}$ response functions. 
In the last definitions $J_{\l} = \frac{1}{S} \int d\bx  \ \bs{j}_{\l}(\bx)$, with $\l=(A,P)$,
and 
$\chi_{J_\l,J_{\l'}}(t) = - i \theta(t) \quave{\left[ J_{\l}(t),J_{\l'} \right]}$.
The pure diamagnetic response $\chi_{AA}$ should not
be confused with the term $\frac{e^2}{m} \rho_0$
in Eq.~\ref{draft_eq:sigma_def}
which instead 
is the diamagnetic contribution
of the probing field that cancels the $\frac{1}{\o}$ divergence 
in the imaginary part of $\s$. 


In order to compute correlation functions 
we make the approximation of decoupling 
electronic and photonic contribution in the diamagnetic term 
on the Hamiltonian (\ref{draft_eq:H_AA}) as
\begin{equation}
\begin{split}
\Psi^\dagger (\bx)  &\Psi(\bx) \bd{A}^2(\bx) 
 \simeq \\
&
\quave{\Psi^\dagger (\bx) \Psi(\bx)}_0 \bd{A}^2(\bx)
+
\quave{\bd{A}^2(\bx) }_0 \Psi^\dagger (\bx) \Psi(\bx) 
\end{split}
\label{draft_eq:a2_decoupling}
\end{equation}
where $\quave{\cdot}_0$ indicates 
averages computed using the non-interacting Hamiltonians.
This decoupling is customarily understood  
in the standard definitions of Dicke-Hopfield models 
for QED.
In particular, the $\bd{A}^2$ term in the Hopfield 
Hamiltonian is obtained by approximating 
the density operator with a number, which is 
equivalent to the decoupling in Eq.~\ref{draft_eq:a2_decoupling}.
As such, this approximation fully retains the effect of the 
$\bd{A}^2$ term of bounding from below 
the photon 
spectrum.~\cite{ciuti_bastard_carusotto,schafer_diamagnetic,nogo_PRL_1975,giacomo_antoine_SXI,andolina_nogo}

\subsection{Current-current response functions}
\label{subsec:jj_response}
With the above decoupling the interaction term
is such that the photonic 
degrees of freedom can be integrated out
to obtain effective electron-electron interactions
mediated by photons, see appendix~\ref{app:response_functions}. 
This fact can be conveniently used to express
all the above correlation functions in term of the 
single particle photon propagator, whose 
retarded component is defined by
\begin{equation}
\begin{split}
&{\cal D}_{\bq q_z \s}^{\bq' q_z' \s'}(t) = -i \theta(t) \quave{[\Phi_{\bq q_z \s}(t),\Phi^\dagger_{\bq' q_z' \s'}]}
\\
&\Phi_{\bq q_z \s} =
\left(
\begin{matrix}
a_{\bq q_z \s}\\
\ac_{-\bq q_z \s}
\end{matrix}
\right).
\end{split}
\label{draft_eq:Dph_spinor}
\end{equation}
In the empty cavity setup the photon field has
full in-plane translational invariance, \ie ${\cal D} \sim \d_{\bq \bq'}$.
In the presence of the two-dimensional material, 
this is reduced to the 
discrete symmetry of the  crystal, namely ${\cal D} \sim \d_{\bq,\bq+\bd{G}}$
with $\bd{G}$ any vector in the reciprocal lattice. 
However, photons of such small 
wavelengths correspond to energies $\hbar c |\bd{G}| \sim 10^3~\mathrm{eV}$
way larger than any reasonable physical energy 
cutoff $\O_\mathrm{ph}$ 
determined by the cavity mirrors.
We therefore restrict ourself to the $\bd{G}=0$ case
and consider, for each wavevector $\bq$, ${\cal D}(\bq)$ 
as a matrix of dimension $2 N_{\bq} \times 2 N_\bq$, with $N_{\bq} = 2 N_z$ 
the number of modes with in-plane momentum $\bq$ and energy 
smaller than the cutoff, where the factor $2$ counts the 
polarizations
and the $2 \times 2$ structure of the 
each block stems for the Nambu representation.

Thanks to Eq.~\ref{draft_eq:a2_decoupling} we incorporate 
the $\bd{A}^2$ term in the definition of the bare photon propagator 
${\cal D}^{-1}_0(\bq,\o) = D^{-1}_0(\bq,\o) - \Pi_{AA} (\bq,\o) $,
being
$D^{-1}_0(\bq,\o) = \d_{q_z q_z'} \d_{\s \s'} 
\left( \o \hat{\tau}_3 - \o_{\bq q_z \s} + i \G_\mathrm{ph} \hat{\tau}_3 \right)$
with $\G_{\mathrm{ph}}$ a small imaginary 
broadening  mimicking dissipation through the mirrors, and 
$\hat{\tau}_3$ the diagonal Pauli matrix.
$\Pi_{AA}(\bq,\o)$ is the self-energy expression 
due to the first term in Eq.~\ref{draft_eq:a2_decoupling} 
and reads
\begin{equation}
\left[ \Pi_{AA}(\bq,\o) \right]_{(q_z \s)}^{(q_z' \s')} = \d_{\s \s'} \frac{e^2}{m} 
\frac{\g_{\bq q_z}}{\sqrt{L}} \frac{\g_{\bq q_z'}}{\sqrt{L}} 
\bs{v}_{\bq q_z \s}^* \cdot \bs{v}_{\bq q'_z \s}
\rho_0 ~I ~I^\dagger
\label{draft_eq:PiAA}
\end{equation} 
In the last expression $I$ is the column vector
$
I=\left( \begin{matrix}
1 \\ 1
\end{matrix}  \right)$, so that Eq.~\ref{draft_eq:PiAA}
represents a $2 \times 2$ matrix in the Nambu space.

Exploiting functional integral identities, see appendix~\ref{app:response_functions},
the correlation functions $\chi_{PP}$ can be related to the 
{\it dressed} photon propagator ${\cal D}$ through 
\begin{equation}
\begin{split}
\left[ {\cal D}_0^{-1}(\bq,\o) 
\right.
&
\left.
{\cal D}(\bq,\o) \right.
\left.
 {\cal D}_0^{-1}(\bq,\o) \right]_{(q_z,\s)}^{(q_z',\s)} - \left[ {\cal D}_0^{-1}(\bq,\o) \right]_{(q_z,\s)}^{ (q_z',\s)}
= \\
&
I~I^\dagger~
\frac{\g_{\bq q_z}}{\sqrt{L}} 
\frac{\g_{\bq q_z'}}{\sqrt{L}} 
\bs{v}_{\bq q_z \s}^{*} 
\cdot
\bs{v}_{\bq q_z' \s} \chi_{PP} (\bq,\o)
\end{split}
\label{draft_eq:chi_el_el}
\end{equation}
which is valid for any pair of modes $(q_z,\s)$ and $(q_z',\s)$.
The correlation function $\chi_{PP}$ is obtained by inverting 
Eq.~\ref{draft_eq:chi_el_el} for any pair of non-orthogonal modes.

The mixed correlation functions are obtained 
using equations of motion technique, 
after applying  the decoupling (\ref{draft_eq:a2_decoupling}) 
to the diamagnetic  current  ${\bs j}_{A} = \frac{e^2}{m} \rho_0(\bx) \bd{A}(\bx)$,	
being $\quave{\bd{A}}_0 = 0.$
The correlation function $\chi_{AP}$ reads
\begin{equation}
\begin{split}
\chi_{AP}(\bq,\o) = 
\sum_{q_z q_z' \s}  
&
\rho_0
\frac{\g_{\bq q_z} }{\sqrt{L}}
\frac{\g_{\bq q'_z}}{\sqrt{L}}
\bs{v}_{\bq q_z \s}
\cdot
\bs{v}^*_{\bq q'_z \s}
\\
&
I^\dagger
 \left[ 
 {\cal D}_0(\bq,\o)
 \right]_{(q_z \s)}^{(q_z' \s)} 
 I
 ~
 \chi_{PP}(\bq,\o)
 \end{split}
 \label{draft_eq:chi_el_dia}
\end{equation}
and $\chi_{PA}$ is obtained 
by computing the  advanced component of~(\ref{draft_eq:chi_el_dia}),
$\chi_{PA}(\bq,\o) = \left[ \chi^{adv}_{AP}(\bq,\o)\right]^* $.
Eventually, the purely diamagnetic contribution reads
\begin{equation}
\begin{split}
\chi_{AA}(\bq,\o) = 
\sum_{q_z q_z' \s} 
\rho_0^2
&
\frac{\g_{\bq q_z \s}}{\sqrt{L}} 
\frac{\g_{\bq q_z' \s}}{\sqrt{L}} 
\bs{v}_{\bq q_z \s} 
\cdot
\bs{v}^*_{\bq q'_z \s} \\
&
I^\dagger
 \left[ 
 {\cal D}(\bq,\o)
 \right]_{(q_z \s)}^{(q_z' \s)} 
 I
\end{split}
 \label{draft_eq:chi_dia_dia}
\end{equation}

\subsection{Photon propagator}
Eqs.~\ref{draft_eq:chi_el_el}-\ref{draft_eq:chi_dia_dia} 
reduce the problem of the computation of the 
response function to the computation of the dressed photon 
propagator $\cal D$.
We compute the photon propagator by treating the 
electron-photon interaction at the level of Gaussian fluctuations. 
This approach corresponds to dressing the photon propagator 
with the bare current-current  response function.
The Dyson equation for the matrix ${\cal D}(\bq,\o)$ reads
\begin{equation}
{\cal D}^{-1}(\bq,\o) = {\cal D}_0^{-1}(\bq,\o) - \Pi_0(\bq,\o)
\end{equation}
with the block components of the self-energies 
\begin{equation}
\left[ \Pi_0 (\bq,\o) \right]_{(q_z \s)}^{(q_z' \s)} =
\frac{\g_q}{\sqrt{V}}
\frac{\g_{q'}}{\sqrt{V}}
\bs{v}_{\bq q_z \s} 
\cdot 
\bs{v}_{\bq q'_z \s} 
\chi^0(\bq,\o)
I I^\dagger.
\label{draft_eq:photon_gaussian}
\end{equation}
$\chi^0(\bq,\o)$ is the current-current response function
for the electronic system computed in the absence 
of the light-matter interaction
\begin{equation}
\chi^0(\bq,t) = -i\theta (t)
\quave{\left[ \bd{J}^0_{-\bq}(t), \bd{J}^0_{\bq} \right]}_0,
\end{equation}
with
\begin{equation}
\bd{J}^{0}_\bq = \frac{e}{m} \bd{P}(\bq) = 
\frac{e}{m}\sum_{\bk \n \n'} \bd{p}_{\bk+\bq \n}^{\bk \n'} 
\cc_{\bk+\bq \n}
\ca_{\bk \n'}
\end{equation}
with 
$
\bd{p}_{\bk+\bq \n}^{\bk \n'}  
= -i \hbar \int d\bx \varphi^*_{\bk + \bq ,\nu} e^{i \bq \bx} \nabla \varphi_{\bk \nu'}(\bx)
$,
being $\varphi_{\bk \nu}(\bx)$ the Bloch functions. 
Computation of the bare response function reduces 
to simple convolutions of single-particle Greens functions.
In Matsubara frequencies this reads
\begin{equation}
\begin{split}
\chi^0(\bq,i\O_n) &= 
\frac{e^2}{m^2} \sum_{\bk \nu \nu'} 
\left| \bd{p}_{\bk+\bq \n}^{\bk \n'} \right|^2 \\
& T \sum_{i\o_n} 
G^0_{\bk+\bq ,\n}(i \o_n+i\O_n) 
G^0_{\bk,\n'}(i \o_n)
\end{split}
\end{equation} 
with $G^0_\bk(i\o_n)= \left( i \o_n - \epsilon_{\bk \nu} \right)^{-1}$
with $i \O_n $ and $i \o_n $ representing, respectively,
bosonic and fermionic Matsubara frequencies.
The retarded component is therefore obtained through
analytical continuation $i\O_n \to \O + i 0^+$.
Eventually, $\chi^0$ is plugged into 
Eq.~(\ref{draft_eq:photon_gaussian})
to get the dressed photon propagator.

\begin{figure}
\includegraphics[width=\columnwidth]{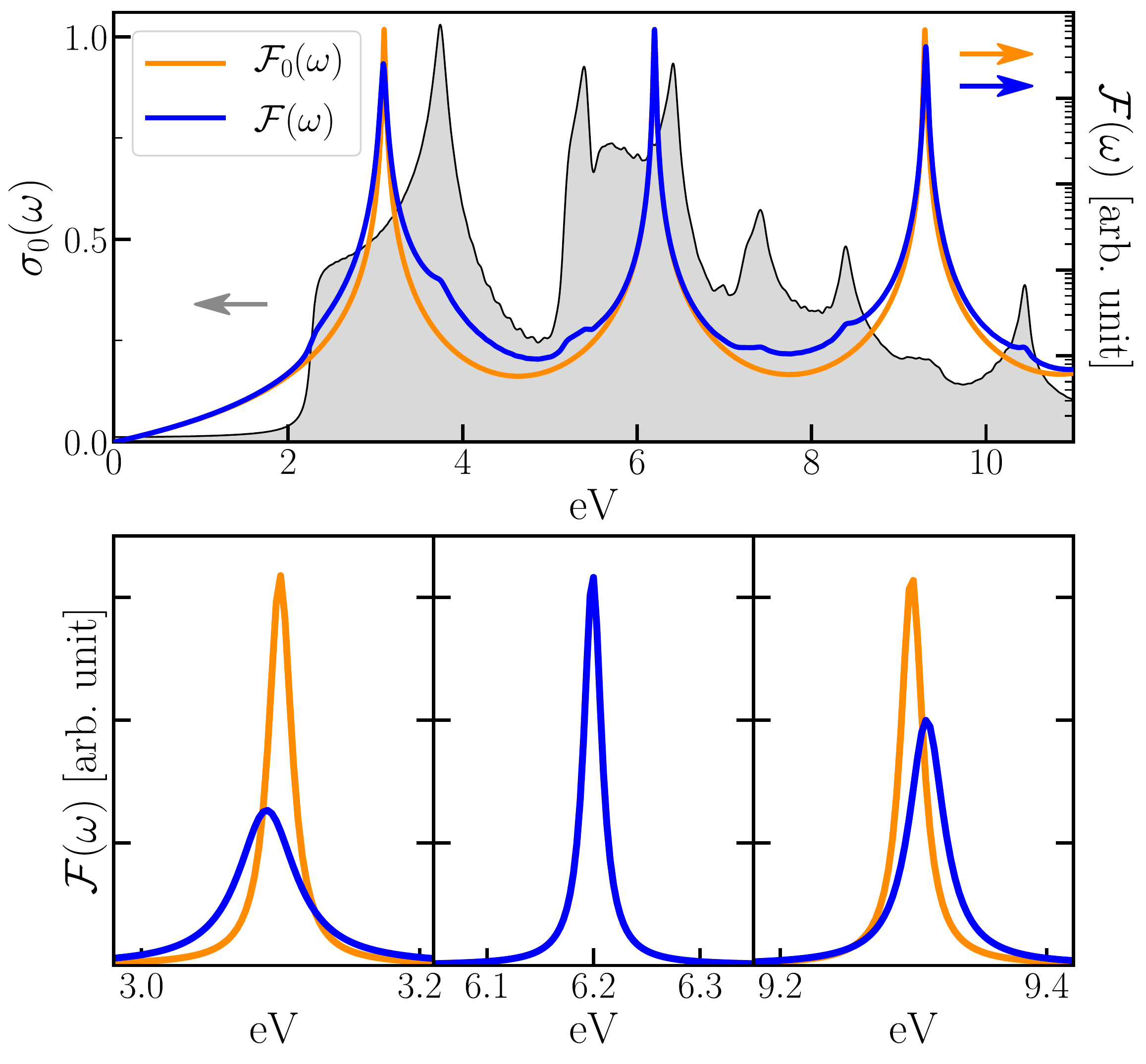}
\caption{
Bare ${\cal F}_0(\o)$
and 
dressed ${\cal F}(\o)$ 
spectral densities for the $\bq=0$ photons for cavity mirrors at a distance $L=0.2~\mu\mathrm{m}$.
Bare photons broadening is set to $\G_\mathrm{ph} = 0.01~\mathrm{eV}$.
Shaded area represents the bare conductivity of the material.
Throughout the paper the conductivity is measured 
with respect to the quantum of conductance $G_0 \equiv \frac{2 e^2}{h}.$
The conductivity is plotted in linear scale, while photon spectra are plotted in logarithmic 
scale (see arrows). 
Bottom panels show the photon spectral densities in linear scale around the single 
photon resonances.
}
\label{fig:fig2}
\end{figure}

\section{Optical dressing of the electronic response}
\label{sec:results_transport}
In this section we present the main results showing the  
effects of the light-matter hybridization in the homogeneous  ($\bq = 0$)
response obtained using the above approximation scheme. 
As it is clear from  Eqs.~\ref{draft_eq:chi_el_el}-\ref{draft_eq:chi_dia_dia}, in the Gaussian
approximation only the 
$\bq=0$ photons contribute to the dressing of the $\bq = 0$ response.
Corrections beyond the Gaussian 
approximation will include interaction with photons at finite 
in-plane momentum and will be discussed in Sec.~\ref{sec:quantum}.
 
\begin{figure}
\includegraphics[width=\columnwidth]{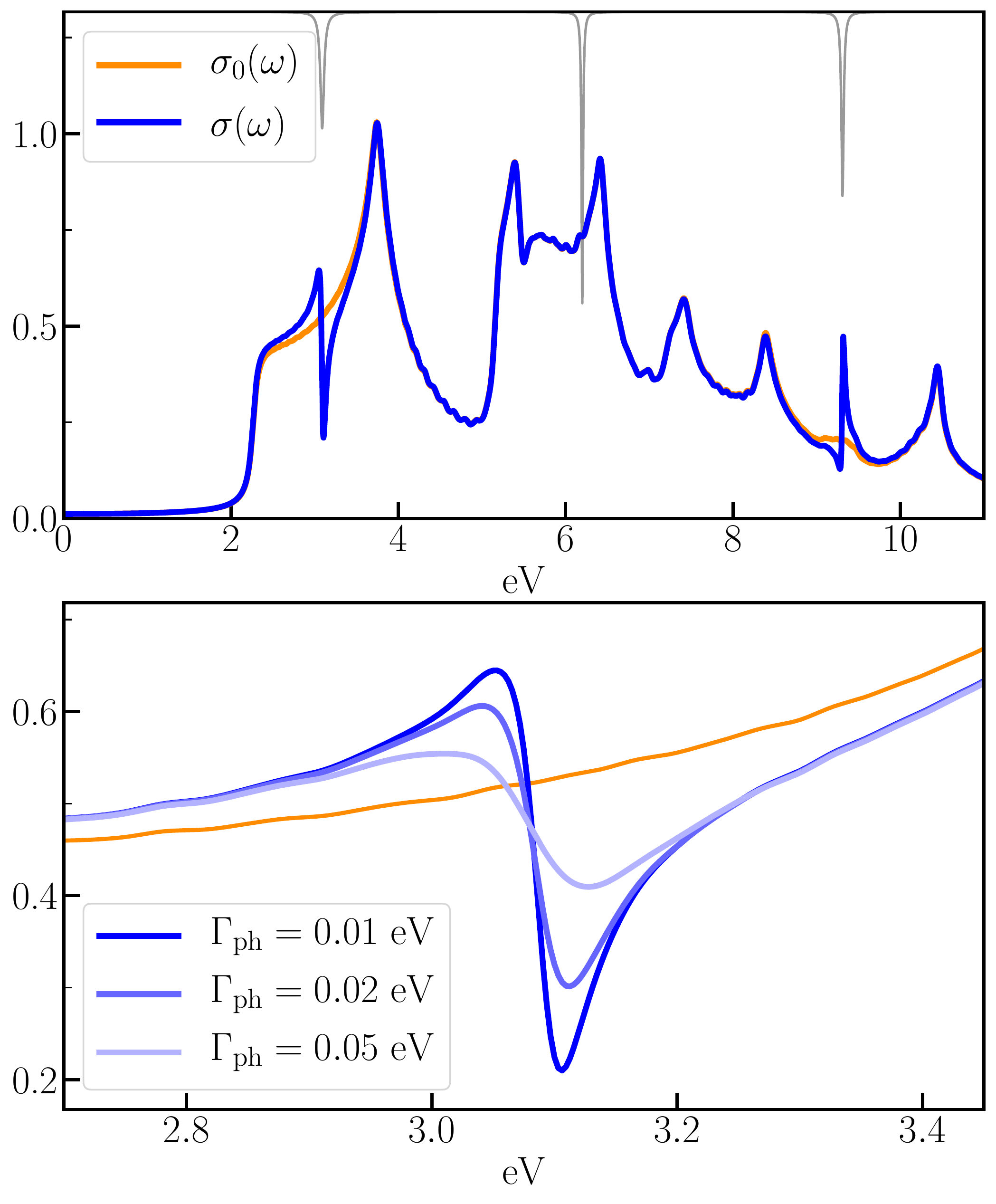}
\caption{
Top: Bare $\s_0$ (orange) and dressed $\s$ (blue) 
conductivities for cavity length $L=0.2~\mu \mathrm{m}$.
The inverted photon spectrum is reported in the top part of the panel.
Photon broadening is $\G_\mathrm{ph} = 0.01~\mathrm{eV}$.
Bottom:
Blow up of the asymmetric feature in the dressed 
conductivity  around $3.1~\mathrm{eV}$ for 
different values of the photon broadening.
}
\label{fig:fig3}
\end{figure}
\begin{figure}
\includegraphics[width=\columnwidth]{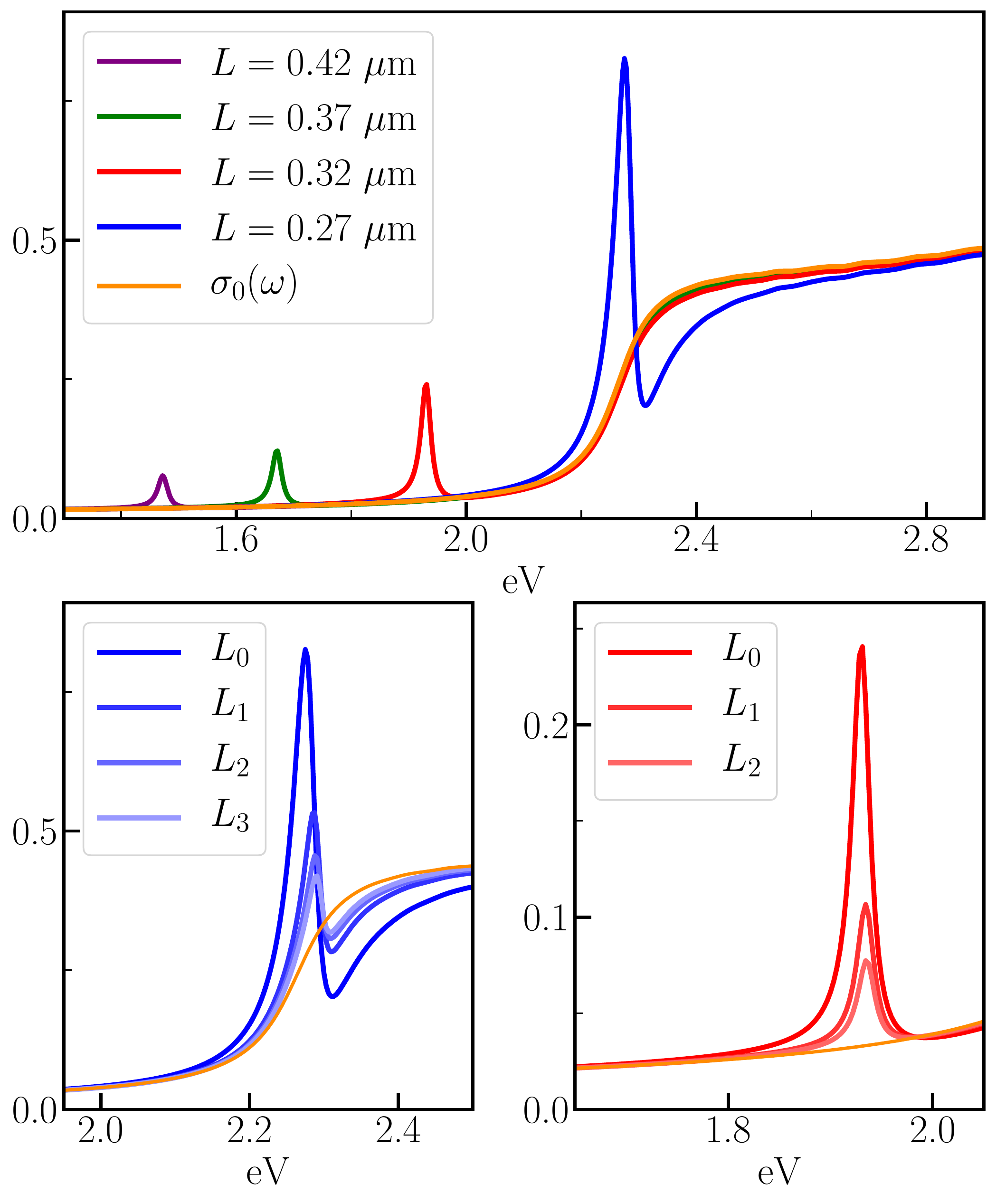}
\caption{Top: 
Dressed conductivity at different cavity lengths.
Bottom: 
Conductivity at resonant lengths $L_n = (2n+1) L_0$ 
for $L_0 = 0.27~\mu\mathrm{m}$ (left) and $L_0=0.32~\mu\mathrm{m}.$  
}
\label{fig:fig4}
\end{figure}

In Fig.~\ref{fig:fig2} we report the total spectral density of 
photons at zero in-plane momentum 
${\cal F}_{\s}(\o) = -\frac{1}{\pi} \sum_{q_z,q_z'} \mathrm{Im} 
\left[ {\cal D}^{11}(\bq=0,\o)\right]_{(q_z \s)}^{(q'_z \s)}$
for a cavity with $L=0.2~\mu\mathrm{m}$.
The symbol ${\cal D}^{11}$ denotes the normal component,
\ie $\quave{a \ac}$, of the Nambu photon propagator.
Due to the symmetry of the problem the spectral density
is polarization independent ${\cal F}_\s (\o) = {\cal F}(\o).$
The bare and interacting spectral densities are superimposed 
to the bare optical conductivity of the system which characterizes  
the absorption of the material in the absence of 
hybridization with vacuum photons.
The dressed photon spectral density is characterized by a shift 
of the resonances and by a broad redistribution 
of spectral weight indicating hybridization with the 
continuum of optical excitations in the material.
Due to the symmetry of the mode functions the dressing 
vanishes in correspondence to the even, \ie $n=2m$, modes
for which $\bs{v}(z=L/2) =0.$

The broad photon dressing  corresponds to sharp features in the conductivity. 
In Fig.~\ref{fig:fig3} we report the dressed  conductivity
$\sigma(\o)$, computed through Eq.~\ref{draft_eq:sigma_def} and
Eqs.~\ref{draft_eq:chi_el_el}-\ref{draft_eq:chi_dia_dia}.
By comparing to the bare conductivity $\sigma_0(\o)$
we observe sharp asymmetric shifts of spectral weight
in correspondence of the dressed photon modes.
This asymmetric shift is reminiscent of a Fano-like 
profile clearly highlighting the hybridization 
of a single photon mode with the continuum 
of electronic excitations.
The amount of redistributed spectral weight depends 
on the broadening of the photon modes $\G_{\mathrm{ph}}$,
decreasing for larger damping.

The resonant frequency of the asymmetric shift 
is modified by acting on the cavity length. 
Fig.~\ref{fig:fig4} reports the cases in which 
the resonances appear, respectively, 
at the gap edge and inside the optical gap.
In the former case, we show a modification
of the optical gap characterized by the formation of a 
sharp peak at the gap edge with sizeable absorption below the bare gap.
When the resonance moves inside the gap, we instead observe the
formation of in-gap absorption peaks. 
In-gap peaks are much weaker with respect to the sharp 
resonances appearing above gap. 
This is understood as, due to the vanishing small 
optical absorption,  light-matter hybridization below gap is progressively 
suppressed.
Experimental investigations along these lines have been carried out in the context of semiconductor microcavities embedding intersubband transitions~\cite{colombelli_exciton_bounds} following the theory in Ref.~\onlinecite{cortese_strong_coupling_ionizing_transition}.

\begin{figure}
\includegraphics[width=\columnwidth]{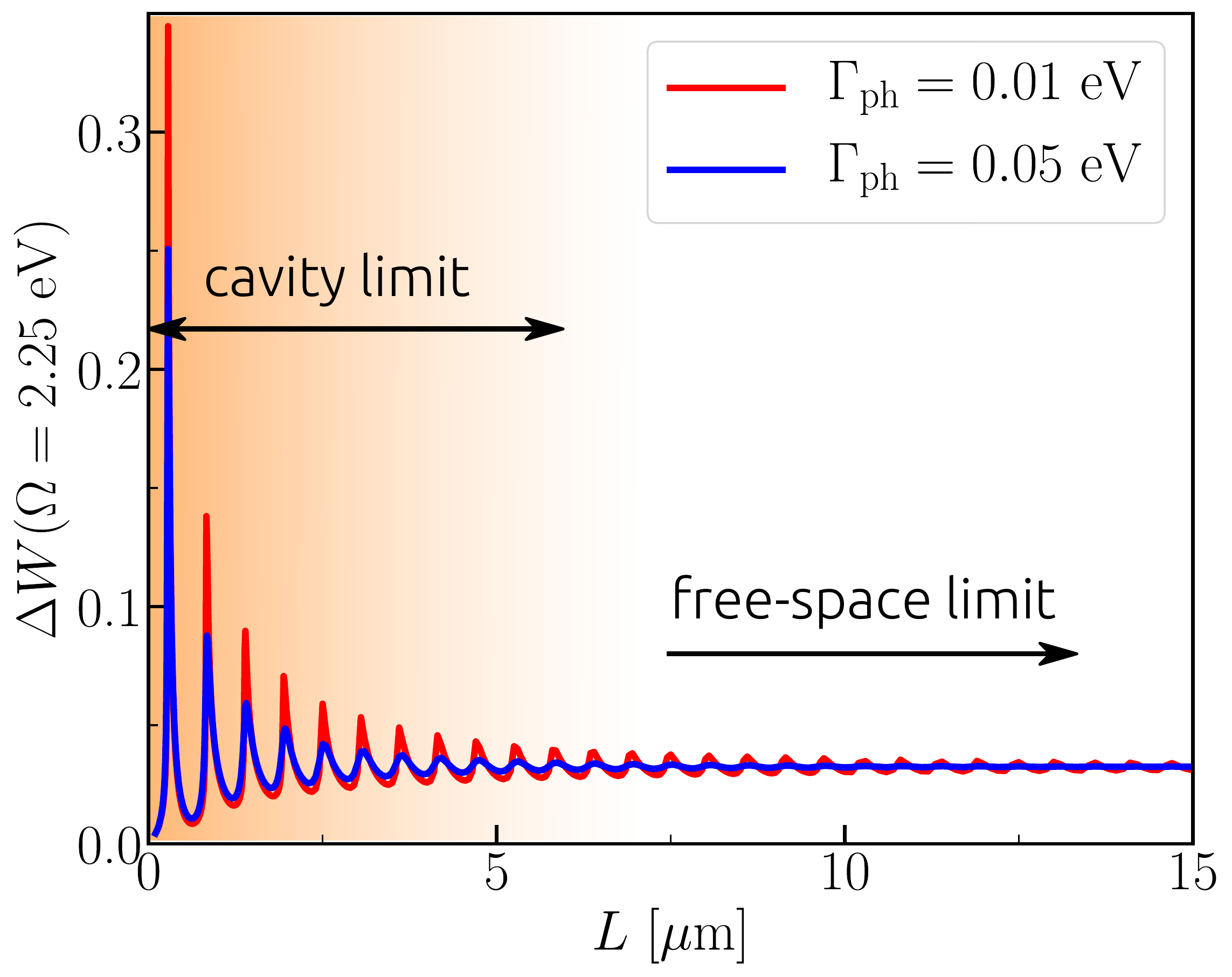}
\caption{
Spectral weight transfer below the gap edge $\O_\mathrm{gap}=2.25~\mathrm{eV}$
as a function of the distance between the mirrors for two values of the photon
damping.
The shaded area highlights the crossover 
from the cavity to the free-space limit.
}
\label{fig:fig5}
\end{figure}

\begin{figure}
\centering\includegraphics[width=\columnwidth]{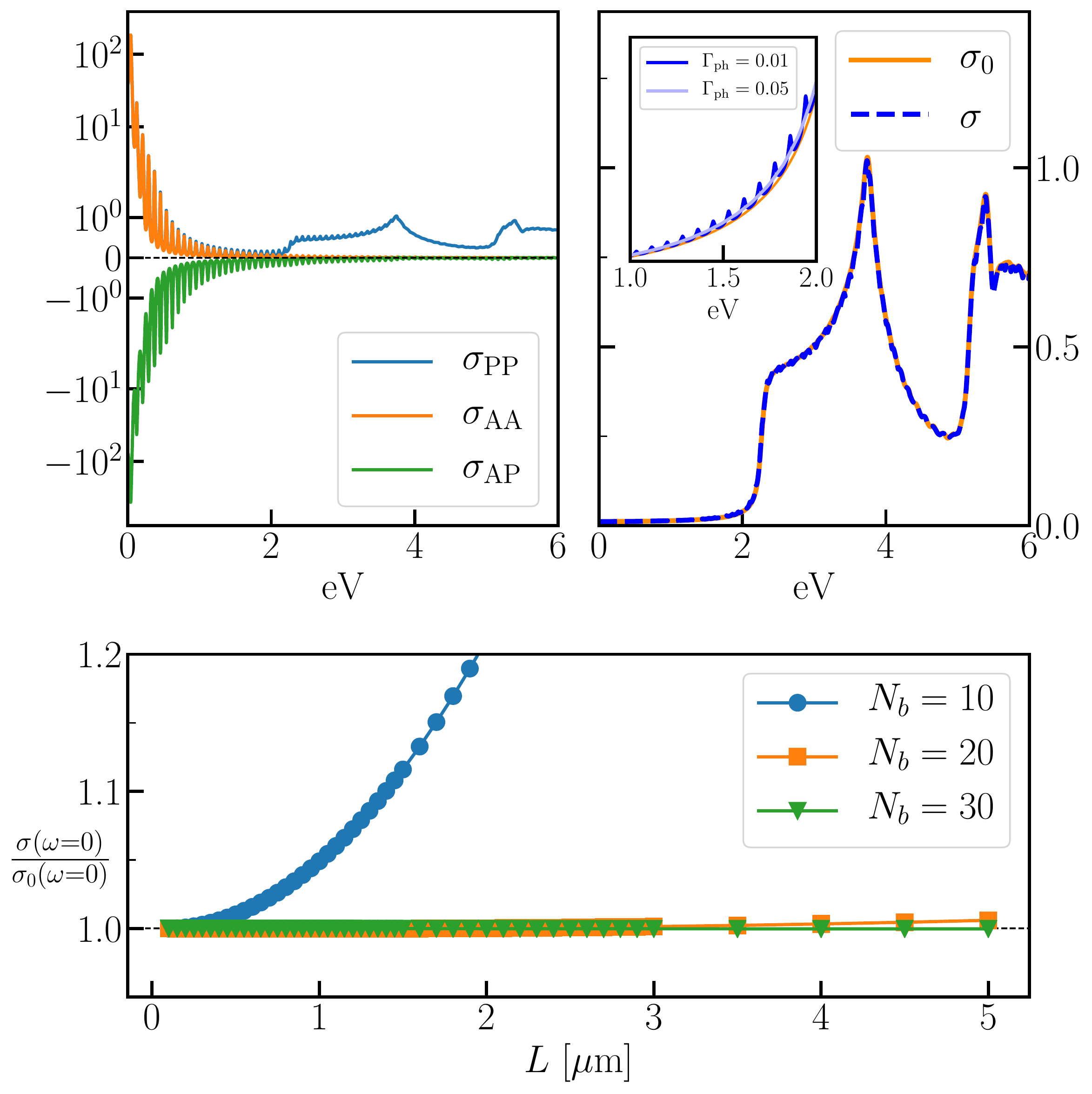}
\caption{
\added{Top panels:} Dressed conductivity in the large 
$L$ limit $L=15~\mu\mathrm{m}$.
Left: purely electronic $\s_{PP}$, purely diamagnetic $\s_{AA}$
and mixed $\s_{AP}$ contributions to the conductivity.
Right: bare conductivity $\s_0$ (orange full) compared to 
the  dressed conductivity 
$\s = \s_{PP} + \s_{AP} + \s_{AA}$ (blue dashed).
Inset: zoom of the dressed conductivity below the gap 
showing the transferred spectral weight
for two values of the photon damping $\G_{\mathrm{ph}}$.
\added{Bottom panel: 
Dressing of conductivity in the static limit as a function of the cavity length,
and for different cutoff in the number of electronic bands $N_b$.}
}
\label{fig:fig6}
\end{figure}

We investigate the effects of confinement on the dressing 
by looking at  cavity lengths $L_n(\O_0)$  featuring 
a resonant mode at frequency $\O_0$. The resonance 
condition is set by 
\begin{equation}
L_n = (2n+1)L_0
\end{equation}
being $L_0 = \hbar c \frac{\pi}{\O_0}$ the length
of a cavity with fundamental mode at the frequency $\O_0.$
The integer $l=2n+1 = \frac{2 L}{\l}$ counts the number 
of half-wavelengths contained in the cavity.
As shown in the bottom panels of Fig.~\ref{fig:fig4}, 
the dressing is suppressed as $n$ is increased.
This fact can be directly correlated to the strength 
of the coupling constants in the Hamiltonian which, 
for a mode at fixed frequency $\O_0$, scales as
$ \g_Q / \sqrt{L} \sim 1/\sqrt{\O_0 L} \sim 1/\sqrt{l} $.

We summarize the effect of light-dressing on the 
conductivity by considering the relative amount of 
spectral weight which is transferred below the gap.
We define the relative variation of spectral weight below 
a given frequency $\O$ as
\begin{equation}
\D W (\O) \equiv \frac{\int_0^{\O} d \o \left[ \s(\o) - \s_0(\o) \right] }{\int_0^{\O} d \o \s_0(\o)}.
\label{draft_eq:deltaW}
\end{equation}
As expected from conductivity sum rules, our results
correctly predict conservation 
of total spectral weight 
$\lim_{\O \to \infty} \D W (\O) \to 0$.
Fig.~\ref{fig:fig5} show the evolution of $\D W(\O=\O_\mathrm{gap})$ computed at
the energy of the bare 
direct optical gap $\O_\mathrm{gap} = 2.25~\mathrm{eV}$, 
as a function of the  distance between the mirrors.
The resonant transfer of spectral weight below the gap
is reflected in the oscillating behavior of $\D W$.
Maxima of transferred spectral weight correspond
to cavity modes resonant with the band edge.
The intensity of the maxima decreases with $L$, as 
anticipated by the results in Fig.~\ref{fig:fig4}.
Eventually for $L \gtrsim 5 \mu\mathrm{m}$  
the amplitude of oscillations  gets significantly 
suppressed and the transferred spectral weight 
smoothly evolves towards an asymptotic value 
independent of $L$.
This evolution clearly highlights the crossover from a 
regime in which the effects of the cavity confinement 
are predominant  to the free-space regime in which 
mirrors becomes irrelevant.

Remarkably, we find that a residual finite dressing, corresponding to 
few percent shift of the spectral weight, persists 
up to the free space limit. 
The dressing in the free-space limit can be 
understood as the regime in which the suppression 
of the couplings due to the larger cavity is 
balanced by the increasing density of 
photon modes.
As physically expected, the free-space 
dressing becomes independent of the photon 
broadening $\G_{\mathrm{ph}}$. In contrast, in the cavity limit 
a larger damping corresponds to a smaller shift 
of the spectral weight.

The effects of the light-matter hybridization in the free-space limit 
can be appreciated by separately looking at the purely 
electronic $\s_{PP}$, purely diamagnetic  $\s_{AA}$
and  mixed $\s_{AP}$ contributions that defines 
the dressed conductivity as $\s(\o) = \s_{PP}(\o) + \s_{AP}(\o) + \s_{AA} (\o)$,
see Eqs.~\ref{draft_eq:chi_el_el}-\ref{draft_eq:chi_dia_dia},
reported in Fig.~\ref{fig:fig6} for $L=15~\mu \mathrm{m}$.
If the couplings with all the photon modes
were set to zero the various contributions would reduce 
to $\s_{AP} = \s_{AA} = 0$ and $\s_{PP} = \s_0$.
However, due to the finite density of modes at low energy
the single contributions shows the $\sim 1/\o$ 
divergence as $\o \to 0$ and the full dressed conductivity 
reduces to $\s \simeq \s_0$ due to the cancellation of terms.
The cancellation is perfect at $\o \to 0$, whereas it leads to 
a residual shift of spectral weight at finite frequency, see inset in Fig.~\ref{fig:fig6}. 

\added{The perfect cancellation of the dressing of the static conductivity 
holds true for all values of cavity lengths, see Fig.~\ref{fig:fig6} bottom panel.
Specifically, we found that the cancellation can be 
highly sensitive to the cutoff in the number of electronic bands 
included in the calculation $N_b$.
Indeed, if $N_b$ is too small, the cancellation of terms 
is not perfect, and a finite dressing of the dc conductivity 
appears for large cavities.
This dressing is fictitious and disappears as soon as the 
electronic cutoff is increased.
The sensibility of static quantities to the truncation 
of the electronic spectrum is a well known feature
of the representation of the light-interaction 
Hamiltonian adopted in this work, Eqs.~\ref{draft_eq:H_AP}-\ref{draft_eq:H_AA}.
Convergence with respect to the truncation of 
the electronic subspace can be understood as 
a the fulfillment of optical sum rules for coupling 
constants.~\cite{debernardis_breakdown,
nori_representations_nat_phys,schuler_deveraux_2021,
jiajun_tight_binding}}

We end this section by observing that 
all the results presented so far 
are converged with respect to the cutoff in the photon 
spectrum.
Specifically, the cutoff was set to 
$\O_{\mathrm{ph}} = 20~\mathrm{eV}$ whereas 
reasonable convergence is already reached for $\O_{\mathrm{ph}} \gtrsim 10~\mathrm{eV}$.

\section{Classical description and quantum corrections}
\label{sec:classical_quantum}
The above results indicate that the 
electronic conductivity contains a 
contribution from vacuum photons 
that is larger the smaller the confining volume 
provided by cavity mirrors.
The effect of light-matter hybridization 
can be generically ascribed  to photon fluctuations in vacuum. 
However, it is possible to give a more practical 
physical description of this dressing.
The conductivity measures the fluctuations of the currents 
in response to an arbitrary small field.
These current fluctuations act as sources of electromagnetic 
fields which, in turn, have a feedback on the current fluctuations.

This observation brings to fundamental question:
to what extent can the optical dressing of the 
electronic response be described in terms of classical 
fields
sourced by current fluctuations in matter?
In this section we address this question by considering 
a classical description of the light-matter hybridization
and comparing it to the quantum description
including corrections beyond the Gaussian approximation
presented in the previous section.

\subsection{Classical description}

\begin{figure}
\centering\includegraphics[width=\columnwidth]{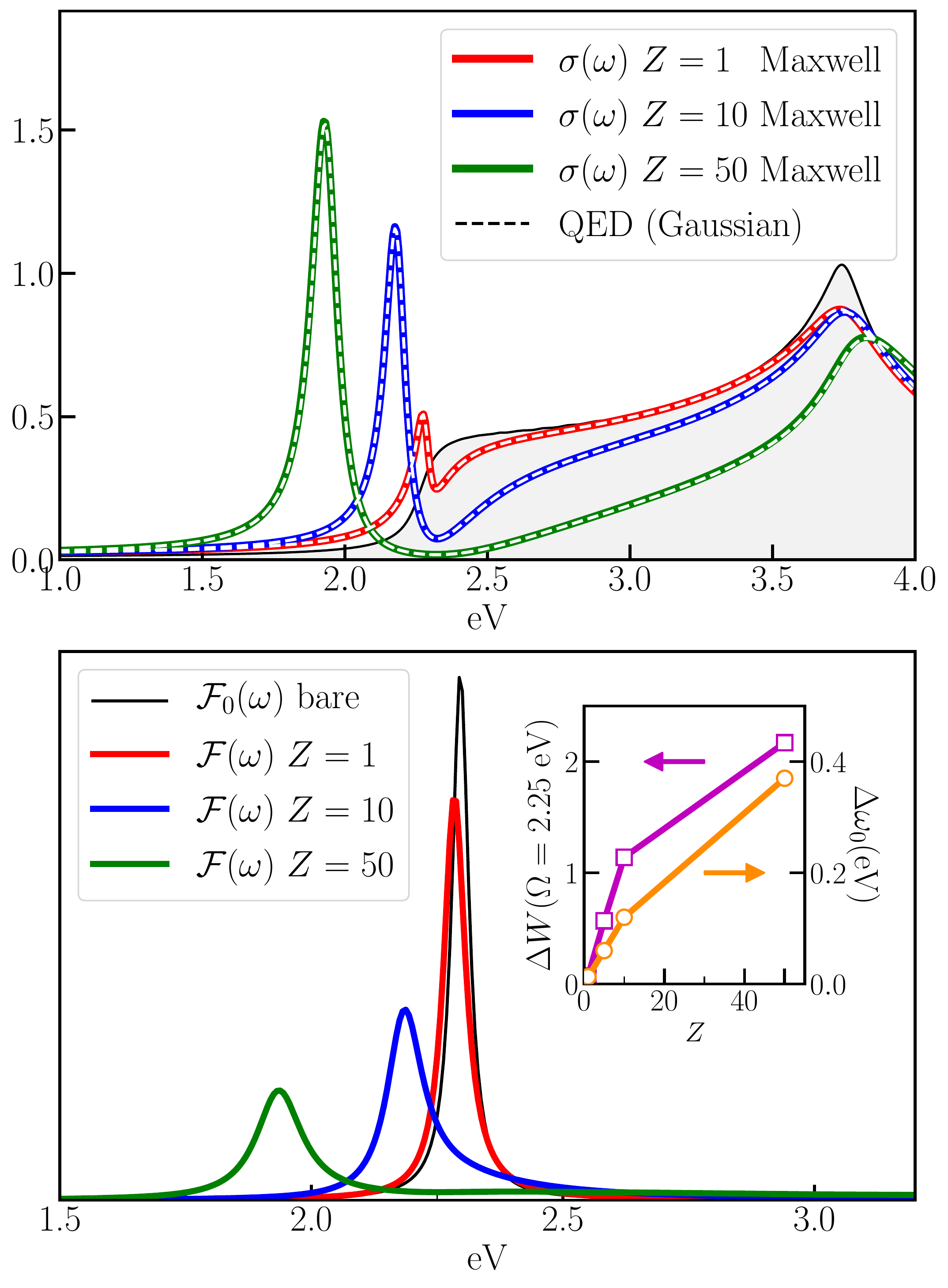}
\caption{Top: Dressed conductivity computed 
using current dressing in Maxwell's equations (full lines) for different
values of the charge renormalization parameter $Z=1$ (red),
$Z=10$ (blue) and $Z=50$ (green). 
White dashed lines indicate the same quantities computed from the 
QED Hamiltonian in the Gaussian approximation for the same values of $Z$.
For comparison the conductivity curves have been scaled by the factor $Z$.
The cavity length is $L=0.27~\mu\mathrm{m}$ and the broadening 
$\G_\mathrm{ph}=0.02~\mathrm{eV}$.
Bottom: 
Photon spectral densities around the cavity fundamental mode 
computed using the QED Hamiltonian in the Gaussian approximation
for the same values of $Z$ shown above.
Insets: shift of the photon resonance of the fundamental cavity mode
as a function of $Z$ (circles, right axis).
$\D \o_0$ is defined as the energy difference 
between the peak of the bare  $\cal F_0$ and 
the peak of the dressed $\cal F$.
Relative shift of the spectral weight Eq.~\ref{draft_eq:deltaW} as a function of $Z$ (squares, left axis).}
\label{fig:fig7}
\end{figure}
The classical description of the conductivity dressing builds on 
the definition of a current in response to an arbitrary small 
applied field.
Calling $J[E_0]$ the current in response to a field $E_0$ 
polarized along $x$ we introduce the decomposition
\begin{equation}
J[E_0] = J_0 [E_0] + \d J [E_0].
\label{draft_eq:current_classical}
\end{equation}
$J_0 = \s_0 E_0$ is the bare current defined by the 
bare conductivity $\s_0$, whereas $\d J$ is the correction 
to the current which is due to the field $E$ sourced by 
the current itself. 
Since $E_0$ is arbitrarily small we can assume that 
also $E$ is small so that we can write $\d J = \s_0 E$.
Eventually, we can define the dressed conductivity as
\begin{equation}
\s(\o) = \lim_{E_0 \to 0} \frac{\partial J[E_0]}{\partial E_0}
= \s_0(\o) 
\left( 1+  
{\eta}(\o) \right)
\label{draft_eq:dressedJ_maxwell}
\end{equation}
with 
$
{\eta}(\o) 
=
\lim_{E_0 \to 0} \frac{\partial E}{\partial E_0}
$
the correction to the bare $\s_0.$

We compute the correction ${\eta}(\o)$
by using Maxwell's equation with $J$ defined in
Eq.~\ref{draft_eq:current_classical} as a source term. 
For the geometry considered in this work
the Maxwell's equations reduce to the 
one dimensional wave-equation
\begin{equation}
\frac{\partial^2 E(z,t)}{\partial z^2} -\frac{1}{c^2} 
\frac{\partial^2 E(z,t)}{\partial t^2} =
\mu_0 \d(z)
\frac{\partial J(t)}{\partial t}
\label{draft_eq:wave_eq}
\end{equation}
with boundary conditions $E(\pm L /2) = 0.$
We expand the field in plane waves and 
transform to the frequency domain.
To compare with the result of the previous section we 
add an imaginary damping $\o \to \o + i \Gamma_\mathrm{ph}$ to match 
the broadening of the photon lines.
Solving the wave-equation for $\frac{E(z,\o)}{E_0}$,
the correction to the conductivity is obtained as 
${\eta}(\o) = \frac{E(z=0,\o)}{E_0(\o)}$.
Close to a resonance $\omega_0$ of the cavity, this gives
\begin{equation}
\sigma(\o) \simeq \sigma_0(\o) \left[
1 - \frac{i\kappa}{\omega - \omega_0 + i(\kappa+\Gamma_{ph})}
\right]
\label{eq:single_mode_maxwell}
\end{equation}
with $\kappa = \mu_0 \sigma_0(\omega_0) / L 
$.
The equation~(\ref{eq:single_mode_maxwell}) describes a shift 
of the spectral weight centered in $\o_0$.
Specifically, the response 
is completely suppressed for $\o = \o_0$ when 
the damping is sent to zero  $\Gamma_\mathrm{ph} \to 0^+$.
The spectral weight is therefore moved to higher (lower)
frequencies depending on the positive (negative) sign 
of the imaginary part of $\sigma_0$.
\added{By taking the static limit 
of Eq.~(\ref{draft_eq:wave_eq}), we also notice that 
the correction trivially vanishes for $\o \to 0$, 
therefore matching the perfect cancellation 
of the dressing of the dc conductivity 
discussed at the end of section~\ref{sec:results_transport}.}

In Fig.~\ref{fig:fig7} we plot the dressed conductivity 
obtained by solving the wave equation~(\ref{draft_eq:dressedJ_maxwell})
for the full multi-mode structure of the cavity.
The dressed conductivity shows remarkable agreement 
with the results of previous section obtained by
applying 
the Gaussian approximation to the QED Hamiltonian.
\replaced{For simplicity, we explicitly show in fig.~\ref{fig:fig7}  
the comparison for the dressing of conductivity only in the limit of small cavity.
However, we found  the same perfect agreement 
also for the dressing in the free-space limit.
}{both in the cavity and the free-space limits
(the latter is not explicitly shown in Fig.~\ref{fig:fig7}).}

To see how the agreement depends on 
the overall coupling strength we introduce a parameter $Z$ 
to renormalize the charge of the electron as $e \to \sqrt{Z} e$
so that the optical absorption of the semiconductor 
increases linearly with $Z$.
As shown in the bottom panel of Fig.~\ref{fig:fig7}, increasing $Z$ 
results in a stronger dressing of the photon spectral density.
Specifically, we see that $Z=50$ produces a huge shift 
$\sim 0.4~\mathrm{eV}$ of the energy of the fundamental mode.
This corresponds to an even larger shift of spectral weight 
in the optical conductivity with a modification of the optical 
gap of about $\sim 0.5~\mathrm{eV}$, clearly indicating 
a regime of strong coupling between light and matter. 

Even in this strong coupling regime the dressing 
of the optical conductivity computed using 
Dyson equation for the photon propagator perfectly 
matches the one computed by using Maxwell's equation. 
We therefore conclude that, independently of the strong coupling 
regime, the classical description fully reproduces results of 
Gaussian approximation for the QED Hamiltonian.

We emphasise that this strong coupling regime, 
signalled by a large shift of the  photon resonance
and spectral weight in the electronic response, 
is equivalent to a large polariton splitting 
due to the hybridization of a single photon mode 
with single matter excitations, 
such as excitons or intersubband transitions.~\cite{mak_qed_2dtmd,ciuti_bastard_carusotto,
liu_MoS2_nat_phot,latini_nano_letters_tmd}
The absence of the double peak structure in the photon 
spectra (Fig.~\ref{fig:fig7}b) is due to the fact 
that in the present case photons hybridize with 
a continuum of electronic transitions giving rise to a smooth 
absorption spectrum.
The results presented can be readily extended to the case 
in which the absorption spectrum is characterized by sharp 
resonances and we do not expect major changes
to the underlying physics.

\subsection{Quantum corrections}
\label{sec:quantum}
The perfect agreement between the classical regime 
and Gaussian treatment of the QED Hamiltonian
can be expected as only the bare response function 
$\chi_0$ enters in the dressing of the photon 
propagator at the Gaussian level
in the same way as only the bare conductivity $\sigma_0$
enters in the classical correction Eq.~\ref{draft_eq:dressedJ_maxwell}.
Therefore, corrections to the classical results must be 
encoded in the corrections beyond the Gaussian approximation 
for the photon propagator.

\begin{figure}
\centering\includegraphics[width=\columnwidth]{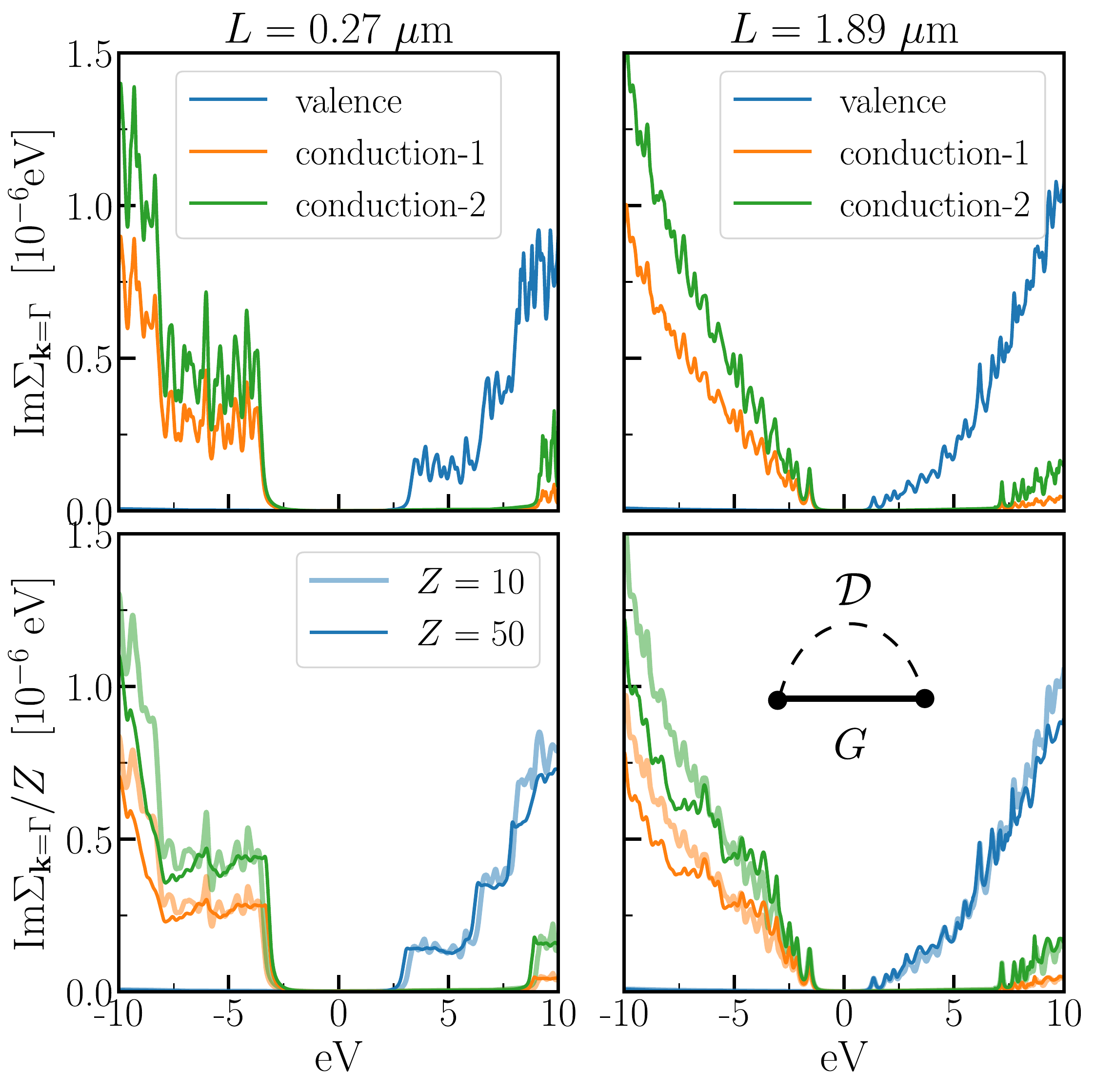}
\caption{
Imaginary parts of the electronic self-energy computed at the 
$\G$-point for the valence and first two conduction 
bands  and for two cavity lengths
$L=0.27~\mu\mathrm{m}$ (left) 
and 
$L=1.89~\mu\mathrm{m}$ (right).
Top panels show the self-energies for $Z=1$.
Bottom panels show self-energies divided by $Z$
for $Z=10$ and $Z=50$.
In the bottom right panel we reported a sketch 
of the self-energy diagram considered.
The full line indicates the electronic Green's function $G$,
while the dashed line indicates the photon propagator ${\cal D}$.
The dots indicate the couplings.
For simplicity, in the sketch we omitted all the 
labels of quantum numbers as well as the energy and momentum conservation at the vertices.
The noise in the self-energies is due to the discretization 
of the $\bq-$grid for momentum integration of
the photon propagator.
}
\label{fig:fig8}	
\end{figure}

In this section we estimate the size of such quantum corrections 
by focussing on the lowest order electronic self-energy corresponding
to the dressing of the electron propagator 
with a single photon line (see sketch in Fig.~\ref{fig:fig8}).
The self-energy expression reads
\begin{equation}
\begin{split}
& [\Sigma (\bk,i \o_n)]_{(\n,\n')}  =
\frac{1}{S} \sum_{\bq} \sum_{\substack{q_z \s \\ q_z' \s'}} \sum_{\m \m'} 
T \sum_{i \O_n} 
K  \times \\
& \times I^\dagger 
\left[ 
{\cal D}(\bq,i \O_n) 
\right]_{(q_z \s)}^{(q'_z \s')}
I \times 
\left[
G(\bk+\bq,i\o_n+i\O_n)
\right]_{\m \m'}.
\end{split}
\label{draft_eq:electron_sigma}
\end{equation}
The constant $K$ depends on all the outer and inner indices and is 
defined  as $K \equiv \frac{\g_{\bq q_z}}{\sqrt{L}} \frac{\g_{\bq q'_z}}{\sqrt{L}}
\left(\bs{v}_{\bq q_z \s} \cdot \bd{p}_{\bk+\bq \n}^{\bk \m} \right)
\left( \bs{v}_{\bq q'_z \s'} \cdot \bd{p}_{\bk+\bq \n'}^{\bk \m'} \right)^*.$
The self-energy corrections are used to dress 
the single-particle electronic Greens function
$G$, which enters in the definition of the self-energy for 
the  photon propagator.
Eventually, a self-consistent procedure is established 
to simultaneously obtain the dressed electronic and photon 
propagators.~\cite{hagenmuller_cavity_transport}


Before computing  the self-energy we estimate the 
order of magnitude of the corrections by considering a 
single electronic transition $\n \to \mu$  
and replacing the photonic and electronic propagators 
in Eq.~\ref{draft_eq:electron_sigma} with the bare ones.
The integration over in-plane momenta 
$\frac{1}{S} \sum_{\bq} \to \int \frac{d\bq}{(2 \pi)^2}$
is restricted at momenta $|\bq| < q_c = \frac{\O_{\mathrm{ph}}}{\hbar c}$.
For cutoff energies 
$\O_\mathrm{ph} \sim 10-20~\mathrm{eV}$ the 
momentum cutoff is $q_c \ll \frac{\pi}{a}$ so that the integration domain
corresponds to a small region around the $\bq=0$ point in the 
Brillouin zone. This fact allows to take the fermionic propagator 
out of the integral  $G_0(\bk+\bq) \simeq G_0(\bk)$.
We define the constant $K_{\n \mu}(\bk) = \frac{\hbar^2 e^2}{2 \epsilon_0 m^2}
|\bd{p}_{\bk \n}^{\bk \mu}|^2$
and take the summation over the frequencies.
We obtain, after analytical  continuation,
$\mathrm{Im} \Sigma_{\nu\to \mu}(\bk,\o) \sim 
K_{\nu \mu}(\bk) \left[ b(\epsilon_{\bk \mu} - \o) + f(\epsilon_{\bk \mu}) \right] 
{\Delta}(\epsilon_{\bk \mu} - \o)$,
with $b$ and $f$ the Bose and Fermi functions,
respectively.
The function $\Delta(\O)$ is 
defined by the photon density of states
$\rho_{\mathrm{ph}}(\O)$ as
$\D(\O) = \frac{\rho_{\mathrm{ph}} (\O) - \rho_\mathrm{ph}(-\O)}{|\O|} $
with 
$\rho_{\mathrm{ph}}(\O) \simeq
\frac{\theta(\O)}{\pi^2} \frac{\O^2}{(\hbar c)^3}.$
In the last approximation 
we considered the density of states in the continuum limit.
At low temperatures, the phase space for the electron-photon 
scattering is 
$b(\epsilon_{\bk \mu} - \o) + f(\epsilon_{\bk \mu}) \sim -\mathrm{sgn}(\o)\theta(|\o| - |\epsilon_{\bk \mu}| )$. 
The largest contributions in the function $\D$
come from the smallest values of 
$|\epsilon_{\bk \mu}|$, so that we estimate the order of 
magnitude of the self-energy by taking 
$\D(\epsilon_{\bk \mu} - \o) \sim \D(-\o)$
from which
$\mathrm{Im} \Sigma_{\nu\to \mu}(\bk,\o) \sim 
K_{\nu \mu}(\bk) \frac{\rho_\mathrm{ph}(|\o|)}{|\o|} $.
By assuming 
$|\bd{p}_{\n}^{\mu}| \sim \frac{\pi}{a}$, 
the constant is  
$K_{\n \m }(\bk) \sim Z \times \mathrm{eV}^3~\mathrm{nm}^3 $
being $Z$ the electronic charge renormalization constant.
On the other hand the photon density of states is
$\rho(|\o|)/|\o| \sim |\o| \times 
10^{-7}~\mathrm{eV}^{-3}~\mathrm{nm}^{-3}.$
We therefore obtain 
$\mathrm{Im} \Sigma_{\n \m} \sim Z |\o| \theta(|\o| - \o_0) \times 10^{-7}~\mathrm{eV}$
being $\o_0$ the fundamental mode of the cavity.

\begin{figure}
\centering\includegraphics[width=\columnwidth]{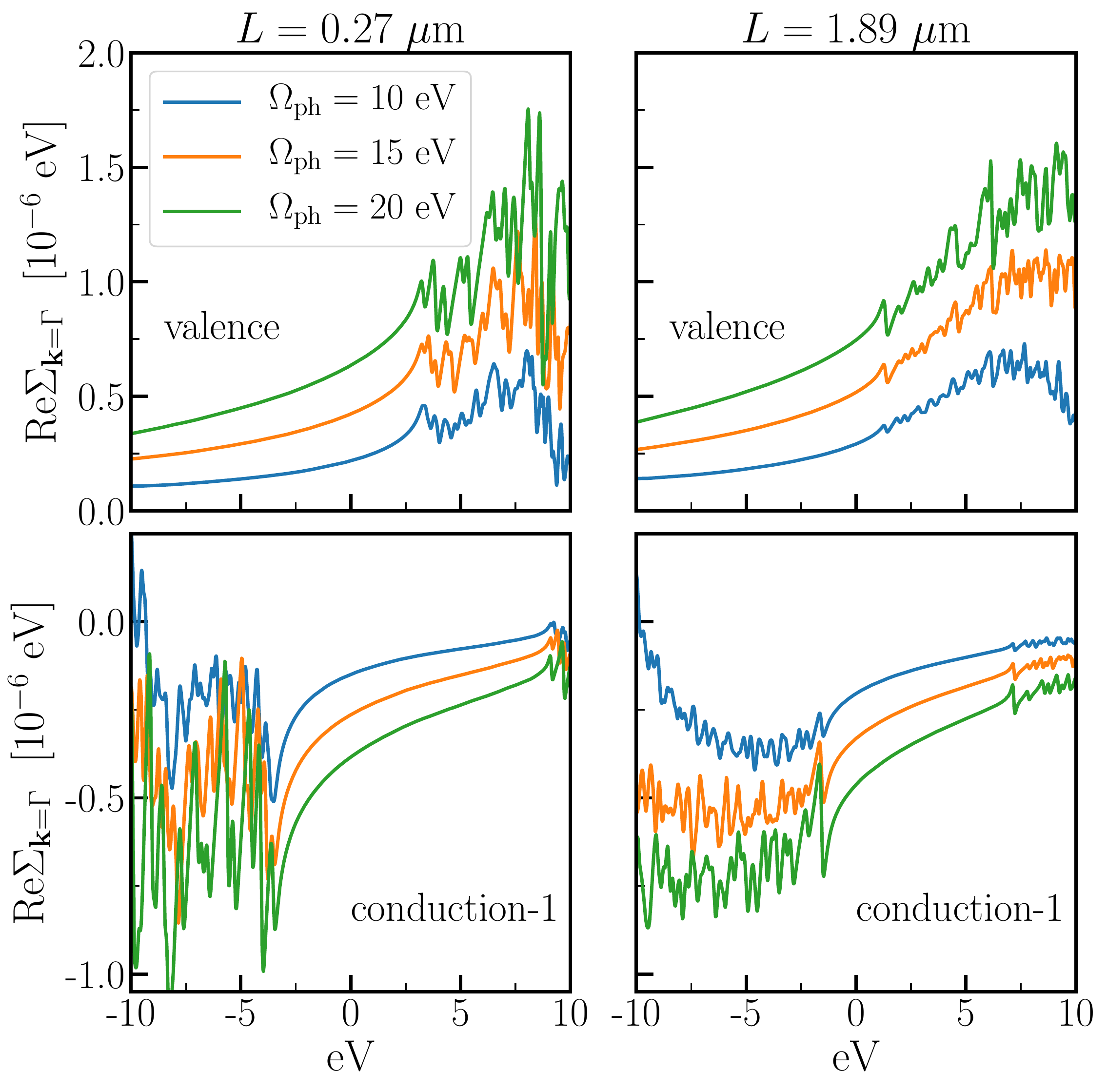}
\caption{
Real parts of the electronic self-energy computed at the 
$\G$-point for the valence (top) and first conduction (bottom)
bands  and for two cavity lengths
$L=0.27~\mu\mathrm{m}$ (left) 
and 
$L=1.89~\mu\mathrm{m}$ (right).
Panels show self-energies for $Z=1$ and different  
cutoff energies $\O_\mathrm{ph} =10~\mathrm{eV}$, 
$\O_\mathrm{ph} =15~\mathrm{eV}$ and 
 $\O_\mathrm{ph} =20~\mathrm{eV}$.
}
\label{fig:fig9}
\end{figure}

The above estimation, valid for a single electronic transition, 
is confirmed by numerically integrating the self-energy using 
the dressed  photon propagator and including all the electronic 
transitions within the cutoff in the number of electronic bands.
Fig.~\ref{fig:fig8} reports the imaginary part of the self-energy 
computed at the $\G$ point for the valence and 
the first two conduction bands. 
We consider two values of cavity lengths characterized  
by modes resonant with the gap edge for which the 
largest modification of the optical gap is observed, see Fig.~\ref{fig:fig4}.
The self-energies are characterized by some noise
which is due to the finite grid used to discretize the 
$\bq-$space for momentum integration of the photon propagator.

$\mathrm{Im} \Sigma$  
vanishes for frequencies smaller than the fundamental mode 
of the cavity and than increases with $|\o|$
remaining of the order 
of  $\lesssim 10^{-6}~\mathrm{eV}$ for $|\o| \lesssim 10~\mathrm{eV}$, 
in agreement with the above estimation
obtained for a single electronic transition.
In the limit of small cavity it is possible 
to appreciate a step-like behaviour due to characteristic jumps 
in the photon density of states. The jumps are washed out for 
larger cavities due to the smoother density of state of photons.
We also show that the self-energy scales linearly with $Z$ 
(bottom panels).
For larger $Z$, the self-energies are less noisy due to the additional broadening 
in the photon spectral density due to the strong coupling dressing.

Fig.~\ref{fig:fig9} reports the real part of the self-energy
which measures 
the renormalization
of the energies of the single-electron excitation 
as given by the pole equation
$\o_* - \e_{\bk \n} - \mathrm{Re} \Sigma_{\n \n}(\bk,\o_*) = 0.$
We observe that the real-part has an intrinsic dependence on 
the photon cutoff $\O_\mathrm{ph}$.
This dependence is readily understood as the real part 
is related to the imaginary part via a Kramers-Kronig  
transformation (KKT) 
$\mathrm{Re} \Sigma(\o) = -\frac{1}{\pi} \int d \o' \frac{\mathrm{Im} \Sigma(\o')}{\o-\o'} $.
Since the imaginary part 
is proportional to the photon density of states 
it increases as a function of frequency
up to the cutoff $\O_\mathrm{ph}$ 
after which it drops to zero.
Therefore  a larger high energy cutoff will results 
in a larger real part at low frequencies.

This observation indicates that 
renormalization of the electronic poles by
self-energy corrections are actually dominated by 
off resonant high-energy modes. 
As the photon density of states diverges at $\o \to \infty$, 
we would expect a divergent real-part for $\O_{\mathrm{ph}} \to \infty.$
We expect this divergence to correspond to the 
well known divergences encountered in the non-relativistic 
description of effects like Lamb shifts.
These divergences are generically cured  by introducing 
a cutoff at the order of the Compton scale.~\cite{bethe_lamb}
As the goal here is to understand the effects of 
the photons confined by the mirrors below the physical 
cutoff of the plasma frequency, we have assumed that all the effects 
of photons at such high-energy scales to be already included in the  
definition of the bare electronic dispersion, see Sec.~\ref{sec:mulit_mode_QED_Hamiltonian}.
As such, we can argue away these divergences and 
restrict ourselves to the interaction effects below 
the physical cutoff $\O_\mathrm{ph}$
and focus on their dependence on the cavity confinement.

For cutoff in the $\O_\mathrm{ph}\sim 10-20~\mathrm{eV}$ 
range the real part of the self-energy remains of the same 
order of magnitude of the imaginary one. 
Even considering a very large  $Z \sim 10^2$ corrections in the 
$\mathrm{eV}$ range are of the order 
$\lesssim 10^{-4}~\mathrm{eV}$,
meaning that, compared to typical electronic 
energy scales, renormalization 
of the pole of the electronic Green's function 
can be considered as negligible.
Most importantly, we observe that the overall magnitude 
of the corrections weakly depends on the cavity confinement
which is consistent with the above arguments
showing that the renormalization mostly depends on the cutoff.

We point out that the smallness of these 
corrections can be generically ascribed to the 
smallness of the photon density states at low energies.
To better appreciate this aspect, it is useful to 
compare the density of states of photons with 
that of acoustic phonons which 
have the same dispersion of photons, 
with speed of sound $c_0 \ll c$, 
and couple 
to electrons in a similar way, 
with different matrix elements.
At variance with photons, phonons have an intrinsic 
low-energy cutoff set by the lattice, whereas the photon 
spectrum is unbounded.  
It is immediate to see that at low-energies 
the phonon density of states is enhanced by the factor 
$ \left( c/c_0 \right)^d \gg 1$ being $d$
the dimensionality of the system.
It follows that, while scattering with phonons 
at low energy  generically leads to a dressing of the electronic 
Green's function,
in the case of photons the largest effects are 
expected to come from off resonant modes at high energies 
for which the density of states becomes large. 
Since the cavity modifies the photon spectrum only 
below the mirrors' plasma frequency, it is conceivable that 
cavity confinement has a small effect on the photon-dressing of the 
single particle electron 
Green's function as shown in Figs.~\ref{fig:fig8}-\ref{fig:fig9}.

Based on these observations we conclude that, for all
practical purposes, dressing of the electronic 
Green's function  due to low-energy photons confined by the cavity 
can be neglected.
We emphasise  that this phenomenology is intrinsically different
from the results in Sec.~\ref{sec:results_transport} 
which are instead fully converged with respect to the cutoff $\O_\mathrm{ph}$
and can be explained without invoking any 
modification of the single-electron excitation energies.

Calculation of higher order corrections, included the 
vertex corrections related to the decoupling of 
the diamagnetic term, Eq.~\ref{draft_eq:a2_decoupling}, 
are beyond the scope of this work.
However, we expect that similar arguments related to the 
small photon density of states at low-energy and to the unbound 
growth of density of states up to relativistic energies to apply also 
to higher-order corrections.

The main consequence of the above observations is 
that corrections to the Gaussian approximation 
used for the computing the photon propagator 
are expected to be small and, in general, weakly 
dependent on the confinement of 
the mirrors.
As a result, having shown that results
in the Gaussian approximation can be fully reproduced 
using the classical description independently of the 
strength of the coupling, we conclude that 
quantum effects due to confinement of 
low-frequency  photons can be considered negligible. 

\section{Conclusions}
In this paper we have studied the modification
of the electronic properties due to light-matter 
hybridization in a planar cavity.
We have focused on the electronic conductivity of a 
two-dimensional material placed in 
between two parallel mirrors that confine the 
electromagnetic field on a length scale $L$.
We have treated the light-matter interaction  
by considering the multi-mode  expansion of the non-relativistic QED 
Hamiltonian including all the photon modes 
below an energy cutoff $\O_\mathrm{ph}$, physically
corresponding to the plasma frequency of the mirrors.

We have shown sharp signatures of the light-matter 
hybridization corresponding to sizeable redistribution 
of spectral weight at resonant frequencies. 
At the gap edge for optical absorption, 
the light-matter hybridization results into a renormalization of the 
optical gap. 
Following the conductivity as a function of 
the distance between the mirrors,	 we have 
described the crossover from the cavity 
to the free-space limit.
In the cavity limit, the confinement gives rise to significantly 
larger shifts of the spectral weight.
In the free-space limit, we observe a residual 
dressing which is independent of the mirrors.

We have shown that both limits can be accurately 
reproduced using a classical description in which
current fluctuations get dressed by self-interaction with 
the fields sourced by the current fluctuations.
By introducing an effective renormalization of the 
electronic charge, we have demonstrated that the 
classical description remains valid even in 
the strong coupling regime, where the 
shift of the cavity resonance due to the light-matter 
interaction becomes of a sizeable fraction of its bare value.

Using this comparison, we have investigated the quantum 
effects of the light-matter interaction by considering 
 corrections beyond the Gaussian approximation.
We have shown that for low-energy photons confined 
in the cavity, these corrections are negligibly small 
up to the very strong coupling regime.
Most importantly, these corrections weakly depend
on the cavity confinement.
As a result, the single-particle \replaced{properties}{excitations}
are not substantially modified by the cavity confinement.

Our results indicate that, despite a significant effect of light-matter 
dressing on the optical gap, the strong coupling regime 
does not automatically correspond to an equally large
modification of the single-particle properties, 
namely the electronic dispersion and the electronic band gap.
\added{This difference can be understood by observing that 
the response functions contributing to the optical conductivity 
have poles at the energies of the so-called bright polaritons, 
which correspond to the transitions from the ground state
to an excited state with the same number of electron upon
emission or absorption of a photon.
On the contrary, the poles in the single-particle Green's 
functions are determined by the virtual transitions
from the ground states to all excited states with one particle 
added or removed. Therefore, unlike the response functions 
the poles in the single particle Green's function 
have no direct connection with the bright or dark polaritons.}

The large modifications of the electronic response  
are mostly captured by the  classical Maxwell equations, 
thus reducing optical dressing due to the low-energy 
photons  confined by the cavity to an essentially classical 
effect.
While the modifications of the electronic response appear in agreement with recent spectroscopic experiments~\cite{colombelli_exciton_bounds},
future work is required to unravel implications of these results 
for different physical situations, such as, for example, the 
response beyond the linear regime, electroluminescence
or problems of quantum 
transport,~\cite{colombelli_APL_2019,
deliberato_ciuti_electron_tunneling,
deliberato_ciuti_electroluminescence,dubail2021_arrowhead,botzung_dark_states}
\added{ and magnetotransport.}~\cite{bartolo_ciuti_magnetotrasport,feist_magnetostransport}

We stress that our results take into account only the effects 
of  fluctuations of the transverse photons, while we have 
neglected effects associated with the longitudinal part of the 
electromagnetic fields for which metallic mirrors can have an 
impact  \eg on the screening of the electron-electron 
interactions.
Eventually, we expect the presented results to apply also 
in the case of  non-metallic mirrors such as dielectric mirrors.

\paragraph*{Note added.} 
Recently, a preprint by J. Li {\it et al.} ~\onlinecite{jianju_multimode} 
appeared which, in a different context, reaches conclusions similar to what 
discussed in Sec.~\ref{sec:quantum}
regarding the smallness of the effects of band renormalization in a planar cavity.

\section{Acknowledgements}
We acknowledge discussions with
Martin Eckstein, Jiajun Li, Jean-Marc Triscone, Yannis Laplace,
Antoine Georges,
Daniele Fausti,
Alexey Kuzmenko,
Daniele De Bernardis, Yanko Todorov, Atac Imamoglu.

This work has been supported by the 	
Swiss National Science Foundation through an
AMBIZIONE grant. 
Part of this work has been supported by the 
European Research Council 
(Gran No. ERC-319286-QMAC).
I.A. acknowledges hospitality at the Collège de France
during the early stages of this project.
I. C. acknowledges financial support from 
the European Union FET-Open Grant 
MIR-BOSE (n.737017), 
the H2020-FETFLAG-2018-2020 
project ”PhoQuS” (n.820392) and from the 
Provincia Autonoma di Trento, partly through the
Q@TN - Quantum Science and Technology in
Trento initiative.

\appendix
\section{Response functions}
\label{app:response_functions}
In this appendix we provides details of the derivation of 
Eqs.~\ref{draft_eq:chi_el_el}-\ref{draft_eq:chi_dia_dia}
for calculations the response functions. 
The starting point is the partition function ${\cal Z}$ written 
as a functional  integral over photonic and electronics degrees of freedom
\begin{equation}
{\cal Z} = \int \prod_Q {\cal D}[\Phi_Q,\Phi^*_Q]
\prod_{\bk \n} {\cal D}[\Psi_{\bk \n},\Psi^*_{\bk \n}]
e^{-{\cal S}[ \Phi,\Phi^*,\Psi,\Psi^*]}
\end{equation}
where $\Phi_Q$ and $\Phi^*_Q$, with $Q\equiv \left( \bq,q_z,\s \right)$, are pairs of conjugate 
complex variables, while $\Psi_{\bk \n}$ and $\Psi^*_{\bk \n}$
are pairs of conjugate Grassmann variables.
In the following we will adopt the compact notation
for the functional differential 
$\int {\cal D}_{\Phi} \equiv \int \prod_Q {\cal D}[\Phi_Q,\Phi^*_Q] $
and 
$\int {\cal D}_{\Psi} \equiv \int \prod_{\bk \n} {\cal D}[\Psi_{\bk \n},\Psi^*_{\bk \n}]
$
The action ${\cal S}$ depends on all the variables and reads
\begin{equation}
{\cal S} = {\cal S}^0_{\mathrm{ph}} 
[ \Phi,\Phi^* ]
+ 
{\cal S}^0_{\mathrm{el}} 
[ \Psi,\Psi^* ]
+ 
{\cal S}_{int}[ \Phi,\Phi^*,\Psi,\Psi^*]
\end{equation}
where ${\cal S}^0_{\mathrm{ph}} = - \int_0^{\b} d\t d\t' \sum_{Q} 
\Phi^*_Q (\t)
\left[ 
D_Q^0  (\t-\t')
\right]^{-1}
\Phi_Q(\t')$
and
$
{\cal S}_{\mathrm{el}}^0=
- \int_0^{\b} d\t d\t' \sum_{\bk \n} 
\Psi^*_{\bk \n} (\t)
\left[ 
{\cal G}^0_{\bk \n}  (\t-\t')
\right]^{-1}
\Psi_{\bk \n}(\t')
$
are the non interacting actions
with ${D}_Q^0$ and ${\cal G}_{\bk \n}^0$ 
the bare photonic and electronic propagators, respectively.
The interaction action is split 
as 
${\cal S}_{int} = {\cal S}_{AP} + {\cal S}_{AA}$
with 
\begin{equation}
{\cal S}_{AP} = \int_0^\b
d\t H_{AP} [\Phi(\t),\Phi(\t)^*,\Psi(\t),\Psi(\t)^* ]
\end{equation}
and 
\begin{equation}
{\cal S}_{AA} = \int_0^\b
d\t H_{AA} [\Phi(\t),\Phi(\t)^*, \Psi(\t),\Psi(\t)^* ]
\end{equation}
with $H_{AP}$ and $H_{AA}$ the Hamiltonian
defined in Eqs.~\ref{draft_eq:H_AP}-\ref{draft_eq:H_AA}.
Thanks to the decoupling Eq.~\ref{draft_eq:a2_decoupling}
the term ${\cal S}_{AA}$ can be absorbed in the non interacting 
photonic action ${\cal S}_{\mathrm{ph}}^0 $ 
upon redefining the photonic propagator 
$D^0 \to {\cal D}_0$ with self-energy in Eq.~\ref{draft_eq:PiAA}. 
Specifically, after introducing the vectorial representation
$
\bs{\Phi}_Q \equiv
\left( 
\begin{matrix}
\Phi_Q \\
\Phi^*_{-Q}
\end{matrix}
\right)
$
we have
\begin{equation}
{\cal S}_{\mathrm{ph}}^0 + {\cal S}_{AA} 
\to 
-\frac{1}{2} 
\int_0^\b
d\t d\t' 
\sum_{Q Q'}
\bs{\Phi}_Q^{\dagger}(\t)  
\left[
{\cal D}_0(\t -\t')
\right]_{Q Q'}^{-1} \bs{\Phi}_{Q'}
\end{equation}
where the photonic propagator is defined as
\begin{equation}
\left[ {\cal D}_0(\t -\t')
\right]_{Q Q'}^{-1} =
-\d(\t-\t')
\left( 
\d_{Q Q'}\overline{\partial}_\t + \overline{\o}_{Q Q'}
\right)
\label{app_eq:D0_inv}
\end{equation}
where $\overline{\o}_{Q Q'}$ are the energy of the modes 
dressed by the $\bd{A}^2$ term via the self-energy
Eq.~\ref{draft_eq:PiAA}
and the symbol
$\overline{\partial}_\t = 
\left(
\begin{matrix}
\partial_\t & 0 \\
0 & -\partial_\t
\end{matrix}
\right).
$
Notice that upon decoupling of the diamagnetic term
there is also a purely electronic term 
that can be reabsorbed in 
the non-interacting electronic Hamiltonian. 
This term enters as a redefinition of the chemical 
potential and will be discarded.

With the above simplifications the interacting Hamiltonian
can be written as
\begin{equation}
{\cal S}_{int}= \frac{1}{2} \int  d \t \sum_Q \bs{\G}^{\dagger}_Q(\t) \bs{\Phi}_Q (\t) 
+ 
\bs{\Phi}^{\dagger}_Q(\t) \bs{\G}_Q(\t)
\end{equation}
where we introduce the spinors
\begin{equation}
\bs{\G}_{Q} 
\equiv
\left(
\begin{matrix}
\G_Q \\ \G^*_{-Q}
\end{matrix}
\right)
=
\frac{\g_{Q}}{\sqrt{L}}
\left( 
\begin{matrix}
1 \\ 1
\end{matrix}
\right)
\bs{v}_{ Q} \cdot \bd{J}_P({-\bq})
\end{equation}
with 
\begin{equation}
\bd{J}_{P}(\bq) = \frac{e}{m} \frac{1}{S} \sum_{\bk \n \n'} \bd{p}_{\bk+\bq \n}^{\bk \n'} \Psi^*_{\bk+\bq \n} \Psi_{\bk \n'}.
\end{equation}
In the compact index notation $-Q \equiv (-\bq q_z \s).$
By integrating over photonic variables 
and by using Gaussian integration we get
\begin{equation}
{\cal Z} = \int {\cal D}_\Psi e^{-{\cal S}_\mathrm{el}
[ \Psi,\Psi^* ]}
{\cal Z}_{\mathrm{ph}}
[\Psi,\Psi^* ]
\end{equation}
with 
\begin{widetext}
\begin{equation}
{\cal Z}_{\mathrm{ph}} 
[ \Psi,\Psi^* ]
\equiv 
\int {\cal D}_\Phi
e^{-{\cal S}^0_{\mathrm{ph}}
-
\frac{1}{2} \int  d \t \sum_Q \bs{\G}^{\dagger}_Q(\t) \bs{\Phi}_Q (\t) 
+ 
\bs{\Phi}^{\dagger}_Q(\t) \bs{\G}_Q(\t)
}
= e^{-\frac{1}{2} \int_0^{\b} d\t d\t' \sum_{Q Q'} \bs{\G}^\dagger_{Q}(\t) 
\left[ \bs{\cal D} (\t-\t') \right]_{Q Q'} 
\bs{\G}_{Q'} (\t') }
\label{app_eq:integrate_photons}
\end{equation}
\end{widetext}
By taking the functional derivative 
$\frac{\d^2}{\d J^i_{P}(-\bq,\t) 
\d J^{j}_P(\bq,\t') }$ of both sides 
of Eq.~\ref{app_eq:integrate_photons}
and by integrating over the fermionic variables
we arrive, after  simple manipulations, at 
Eq.~\ref{draft_eq:chi_el_el} in the main text
with $\chi_{PP}^{ij}(\bq,\t-\t') = \d_{ij} \chi_{PP}(\bq,\t-\t') 
= -\quave{T_\t J^i_P(-\bq, \t) J^{i}_P(\bq, \t')}.$

The correlation functions involving the diamagnetic 
contribution to the current are obtained by first applying 
the diamagnetic decoupling to the diamagnetic current
$\bd{J}_A = \frac{e^2}{m} \int d\bx \rho(\bx) \bd{A}(\bx) $,
with $\rho_0(\bx)$ the electronic density of the non-interacting 
system. 
The mixed correlation function $\chi_{AP}$ is therefore obtained by using 
the equations of motion.
Specifically, it is straightforward to check that  
\begin{equation}
-\overline{\partial_\t} \bs{\Phi}_Q(\t) = \sum_{Q'} \overline{\o}_{Q Q'} \Phi_{Q'}(\t) + 
\bs{\G}_Q
\end{equation}
where in the last equation
the spinors $\bs{\Phi}_Q(\t)$
and $\bs{\G}_Q$ should be considered at the 
operatorial level.
By applying the equation of motion to 
the correlator $\chi(\bq,\t-\t') = -\quave{T_\t J_{A}(-\bq,\t) J_P(\bq,\t')} $
and using the definition of the photon
propagator dressed by the $\bd{A}^2$ term,
Eq.~\ref{app_eq:D0_inv},
we arrive at Eq.~\ref{draft_eq:chi_el_dia} in the text.

Eventually the purely diamagnetic response $\chi_{AA}$
is straightforwardly written in term of the photon propagator 
by direct expansion of the vector potential operator.

We notice that in all the above steps we have always assumed 
that the photon propagator keeps the full in-plane translational 
invariance ${\cal D} \sim \d_{\bq \bq'}$ as discussed in
Sec.~\ref{subsec:jj_response}.

\bibliography{biblio_cavities}

\end{document}